\documentclass[a4paper,11pt]{article}
\usepackage{jheppub} 
\usepackage[normalem]{ulem}
\usepackage{xcolor}
\usepackage{colortbl}
\usepackage{multirow}
\usepackage{slashed}
\usepackage{dsfont}
\usepackage{caption}
\usepackage{subcaption}
\usepackage{comment}
\usepackage{float}

\newcommand{\be}{\begin{eqnarray}}
\newcommand{\ee}{\end{eqnarray}}
\newcommand{\bea}{\begin{eqnarray}}
\newcommand{\eea}{\end{eqnarray}}
\newcommand{\nn}{\nonumber}

\newcommand{\beq}{\begin{equation}}
\newcommand{\eeq}{\end{equation}}

\newcommand{\MS}{\widetilde M_S}

\definecolor{darkblue}{rgb}{0.2,0.2,0.9}
\definecolor{colorRTD}{rgb}{.2,.2,.7}

\definecolor{colorHD}{rgb}{.2,0.9,.0.9}

\title{Landscapes at Colliders}

\author[1,2]{Raffaele Tito D'Agnolo,}
\affiliation[1]{Universit\'e Paris-Saclay, CEA, CNRS, Institut de Physique Th\'eorique, 91191, Gif-sur-Yvette, France}
\affiliation[2]{Laboratoire de Physique de l'\'Ecole Normale Sup\'erieure, ENS, Universit\'e PSL, CNRS, Sorbonne Universit\'e, Universit\'e Paris Cit\'e, F-75005 Paris, France}
\author[1]{Manuel Ettengruber,}
\author[3,4,5]{Lian-Tao Wang}
\affiliation[3]{Department of Physics, University of Chicago, Chicago, IL 60637, USA}
\affiliation[4]{Enrico Fermi Institute, Kavli Institute for Cosmological Physics, and Leiweber Institute for Theoretical Physics, University of Chicago, Chicago, IL 60637, USA}
\affiliation[5]{HEP Division, Argonne National Laboratory,
9700 Cass Ave., Argonne, IL 60439, U.S.A.}

\emailAdd{raffaele-tito.dagnolo@ipht.fr}
\emailAdd{manuel-meinrad.ettengruber@ipht.fr}
\emailAdd{liantaow@uchicago.edu}

\abstract{Theories with a large number of long-lived metastable vacua are our only concrete explanation for the puzzling value of the Cosmological Constant (CC). The energy scales where these vacua are realized are unknown. In this work, we consider the possibility that a sector of this landscape of vacua is within experimental reach and discuss its signatures at colliders. We find that striking large-multiplicity final states might have gone undetected due to their relatively small total energy. In particular, this could lead to new  exotic Higgs decays, which are both intriguing and challenging to search for. In addition to a general phenomenological analysis of these theories, we also discuss an explicit model where the small values of the CC and the Higgs mass are jointly explained by Weinberg's anthropic argument and a low energy landscape.
}

\begin{document}

\maketitle

\flushbottom

\section{Introduction}
The two most relevant operators of the Standard Model have coefficients that appear mysteriously fine-tuned. The vacuum energy $\Lambda$ and the Higgs mass $m_h^2$ are orders of magnitude below the apparent cutoff of the theory, the Planck mass $M_{\rm Pl}$.

Explaining dynamically the ratio $m_h^2/M_{\rm Pl}^2 \simeq 10^{-34}$ has driven theoretical particle physics for the past decades, leading to models based either on supersymmetry~\cite{Dimopoulos:1981zb, Nilles:1983ge, Haber:1984rc, Martin:1997ns} or scale invariance~\cite{Kaplan:1983fs, Kaplan:1983sm, Georgi:1984ef, Dugan:1984hq, Giudice:2007fh, Arkani-Hamed:2001nha, Arkani-Hamed:2002sdy, Arkani-Hamed:2002ikv, Chacko:2005pe, Burdman:2006tz, Craig:2015pha} and in some cases on both~\cite{Gherghetta:2003he, Sundrum:2009gv, Redi:2010yv, Gherghetta:2011wc, Caracciolo:2012je}. The absence of new physics at LEP, at the LHC, and in the tests of the approximate symmetries of the Standard Model (flavor, CP, baryon and lepton numbers) is putting some strain on these paradigms that naively predict new physics below $\simeq 4\pi m_h/y_t \simeq$~TeV with generic flavor and CP structures.

In contrast, there is no natural explanation for the ratio $\Lambda/M_{\rm Pl}^4 \simeq 10^{-120}$. The hope of a dynamical mechanism from string theory that sets $\Lambda$ to zero has been challenged by the measurement of the accelerated expansion of the universe in the late nineties~\cite{SupernovaSearchTeam:1998fmf}. This same measurement revived a prediction made by Weinberg ten years earlier~\cite{Weinberg:1987dv} which agrees with the observed order of magnitude of the vacuum energy.

Weinberg's argument is based on the existence of multiple vacua that can be populated in a Multiverse~\cite{Hogan:1999wh, Rees:2004av, Barrow:1986nmg}. If the fundamental theory of nature contains an enormous number of vacua that differ only by the value of $\Lambda$, we can explain our observed universe as the one that supports the existence of macroscopic observers. This hypothesis has gained momentum after the discovery of the vast ``landscape" of vacua of string theory~\cite{Bousso:2000xa, Giddings:2001yu, Maloney:2002rr, Kachru:2003aw, Susskind:2003kw, Douglas:2003um}\footnote{Some of the constructions in this list are now debated and so is the existence of stable dS minima~\cite{Garg:2018reu, Ooguri:2018wrx}, but metastable dS minima are sufficient for the existence of a Multiverse, which has been shown to exist also in more extreme circumstances~\cite{DAgnolo:2024jfu} (i.e. no dS minima at all and no flat potentials, in the sense of~\cite{Ooguri:2018wrx}).}.

After decades of null results at particle colliders, a number of variations of this idea have also been applied to the unnatural value of $m_h^2$~\cite{Graham:2015cka, Csaki:2020zqz, TitoDAgnolo:2021nhd, TitoDAgnolo:2021pjo, Agrawal:1997gf, Dvali:2003br, Dvali:2004tma, Arkani-Hamed:2004ymt, Espinosa:2015eda, Arvanitaki:2016xds, Geller:2018xvz,  Cheung:2018xnu, Giudice:2019iwl, Arkani-Hamed:2020yna, Strumia:2020bdy, Giudice:2021viw, Khoury:2021zao, Chatrchyan:2022pcb, Trifinopoulos:2022tfx, Csaki:2022zbc, Matsedonskyi:2023tca, Hook:2023yzd, Chattopadhyay:2024rha, Benevedes:2025qwt}. If a landscape exists because of string theory and constitutes our only concrete explanation for $\Lambda$, it is natural to ask if it can also explain the Higgs mass.

Proving the existence of a landscape would force a radical epistemological shift in fundamental physics. Some of the phenomena that we regard as fundamental might be just an accident of the special vacuum that we live in. This is a change of perspective not dissimilar to the realization that there is nothing special about our solar system, but we are just one of exponentially many planetary systems distributed throughout a vast, empty universe. In his 1596 {\it Mysterium Cosmographicum}, Kepler explained the orbits of the known planets in terms of Platonic solids. A construction as elegant as it appears futile today. The same fate might very well befall our, equally elegant, dynamical explanations for $m_h^2$.

Therefore, it is important to ask if this vast landscape of vacua can leave any trace that we can detect experimentally. Neither Weinberg's argument nor the ideas that explain $m_h^2$ are anchored to a specific energy scale for the landscape. A subset of these vacua can live at or below the weak scale. If they play a role in the fine-tuning of $m_h^2$ they are coupled to the Higgs and leave observable signatures at the LHC. Additionally, the existence of a large number of new particles at energies parametrically lower than $M_{\rm Pl}$ can be motivated in several ways, including dark matter, baryogenesis, neutrino masses, and the electroweak hierarchy problem ~\cite{Cohen:2018cnq, Dienes:2018yoq, Glioti:2018roy, Dienes:2019krh, Arkani-Hamed:1998wuz, Ettengruber:2022pxf, Ettengruber:2025usk,Dvali:2019ewm, Dvali:2007hz, Dvali:2007wp, Dvali:2007iv, Dvali:2009ne, Dvali:2009fw, Arkani-Hamed:1998jmv, Arkani-Hamed:2016rle}.

In this work, we write down concrete models for the landscape and explore their experimental signatures at colliders. We consider separately a generic model of landscape scalars coupled to the Higgs and a more detailed construction where a low energy landscape can give a joint explanation of $\Lambda$ and $m_h^2$.

We find that a low energy landscape generically predicts long cascades at the LHC, leading to high-multiplicity final states. In vast regions of the parameter space, the total transverse energy of these configurations is too small to have been detected. A particularly interesting benchmark for the production of our landscape scalars is through Higgs decays. The signal tends to spread among many different final states with different particle multiplicities, each with a relatively small branching ratio. Hence, a more inclusive approach, compared to traditional strategies, is necessary. This motivates looking for high-multiplicity, but low-energy, final states, providing a justification for the data-scouting efforts that aim at lowering trigger thresholds~\cite{CMS:2024zhe, Boveia:2703715, Benson:2019752}. For other theoretical proposals in this direction see~\cite{Mariotti:2017vtv, CidVidal:2018eel, CidVidal:2018blh, Knapen:2021elo, Dienes:2011ja, Dienes:2011sa, Dienes:2016kgc}.

The paper is organized as follows. In Section~\ref{sec:generic} we introduce generic quantum field theory models of the landscape. In Section~\ref{sec:joint} we write down a specific model that can solve the CC and electroweak hierarchy problem using only Weinberg's argument. In Section~\ref{sec:pheno} we discuss the phenomenology of these landscapes at colliders. We conclude in Section~\ref{sec:conclusion}.

\section{Field Theory Models of the Landscape}\label{sec:generic}
We are going to approach the problem of building a landscape constructively. We start from the simplest possible setup and then adjust it as we discover non-trivial issues. This pedagogical approach follows for the most part the discussion in~\cite{Arkani-Hamed:2005zuc} and it is useful to better understand the logic behind the final models. 

The first attempt at building a field theory landscape is almost trivial. We can write a generic quartic potential for a scalar $V(\phi)$, without any $\phi\to -\phi$ symmetry, so also linear and cubic terms are present. If $m_{\phi}^2 < 0$ there are two minima at $\langle \phi_i \rangle=\phi_\pm$, with corresponding vacuum energy $V_\pm = V(\phi_\pm)$. We choose $V_+ > V_-$. We take $\phi_\pm$ to be $\mathcal{O}(M_*)$, i.e. the cutoff of the theory, and $m_\phi^2 \sim \epsilon^2 M_*^2$ with $\epsilon \lesssim 1$. We take $\epsilon$ to be the only source of breaking of the shift symmetry of $\phi$, with potential of the form $\epsilon^2 f(\phi)$, so that $\epsilon \ll 1$ is technically natural. If we repeat this construction $N$ times for $N$ decoupled scalars $V_{\rm tot}(\vec \phi)= \sum_{i=1}^N V_i(\phi_i)$ we have built a vast landscape of vacua.

This landscape populates $2^N$ values of the CC, but most of them are around $N \epsilon^2 M_*^4$ rather than zero. The width of the CC distribution is much smaller than its mean and given by $\sqrt{N} \epsilon^2 M_*^4$. To see this, we can write the vacuum energy of a single scalar as
\be
V(\phi_\pm) = \frac{V_+ + V_-}{2}+\xi_\pm \frac{V_+ - V_-}{2}\, ,
\ee
where $\xi_\pm = \pm 1$. The total cosmological constant, summed over all the scalars, is
\be
\Lambda_{\boldsymbol{\xi}} &=& N \overline{V}+ \sum_{i=1}^N \xi_i \left(\frac{V_{i+} - V_{i-}}{2}\right) \, , \nn \\
\overline{V}&\equiv&\frac{1}{N}\sum_{i=1}^N\frac{V_{i+} + V_{i-}}{2}\, .
\label{eq:CCsum}
\ee
In the absence of extra symmetries the average vacuum energy and the difference between the two minima are of the same order $(V_+ + V_-) \sim (V_+ - V_-) \sim \epsilon^2 M_*^4$. The $\xi_i$'s are either $+1$ or $-1$ so the typical value of the sum in the first line of Eq.~\eqref{eq:CCsum} is $\sim \sqrt{N}$. As a consequence this landscape scans the CC by an amount $\sim \sqrt{N} \epsilon^2 M_*^4$ around $\sim N \epsilon^2 M_*^4$. For a more detailed argument based on the central limit theorem see~\cite{Arkani-Hamed:2005zuc}.

If we want to scan the CC around zero we can introduce supersymmetry and a discrete symmetry that forces the superpotential to be an odd polynomial of the $\phi$'s
\be
W_i= \lambda \phi_i^3 - m^2 \phi_i\, . \label{eq:Wphi}
\ee
This can be accomplished by a $Z_4$ R symmetry, as proposed in~\cite{Arkani-Hamed:2005zuc}, that acts as $\phi(x, \theta) \to - \phi(x, i \theta)$.  From this superpotential, we have $\phi_{\pm} = \pm m/\sqrt{\lambda}$. Denote  $W_{i\pm}= W(\phi_{i\pm})$. We have $W_{i+}= -W_{i-}$. Once we turn on gravity, the vacuum energy can scan in the range
\be
- \frac{3 N}{4}\frac{|W_+-W_-|^2}{M_{\rm Pl}^2} \lesssim \Lambda \leq 0 \, . \label{eq:SUSYCC}
\ee
What we are showing is a lower bound coming from the typical value of the width of the CC distribution. Note that for simplicity we took all $W_{i\pm}$ to be equal in Eq.~\eqref{eq:SUSYCC}, i.e. $W_{i\pm}=W_\pm$. One can understand the lower bound in Eq.~\eqref{eq:SUSYCC} by looking at Eq.~\eqref{eq:CCsum}, where the term proportional to $V_+-V_-$ is replaced by $3\sum_i \xi_i^2 |W_{i+}-W_{i-}|^2/4M_{\rm Pl}^2$. 
To explain the observed CC, we need SUSY breaking to occur at a scale $\MS$ parametrically smaller than $W_\pm^{1/3} \sim m/\sqrt{\lambda}$, but if this is the case our landscape finely scans the CC around zero.

If $H_{u,d}$ are neutral under the $Z_4$ R-symmetry, this landscape also scans the $\mu$ parameter of the Higgs potential, 
\be
W \supset \sum_{i=1}^N g_i M_* P(\phi_i/M_*) H_u H_d\, , \quad P(x)= a_1 x+ a_2 x^3+... \label{eq:Wmu}
\ee
The discrete R-symmetry makes $P$ an odd function of the fields, so the effective $\mu$ parameter in the Lagrangian is scanned around zero. Additionally, if the SM fermion superfields transform as $f_a f^c_b(y, \theta) \to - f_a f^c_b(y, i\theta)$ under the same R-symmetry, the Yukawa couplings are given by 
\be
W \supset \sum_{i=1}^N \lambda_{abi} G(\phi_i/M_*) f_a f^c_b H_{u} +... \, , \quad G(x)= b_1 x^2+ b_2 x^4+...
\ee
where $G$ is an even polynomial. Therefore the couplings are not appreciably scanned $\delta Y/Y \lesssim 1/\sqrt{N}$, as was the case for the CC in the non-supersymmetric landscape. This is exactly what we need if we want to invoke Weinberg's anthropic principle or the atomic principle for the Higgs mass~\cite{Arkani-Hamed:2005zuc}. If only dimensionful SM parmeters scan in the landscape, but the dimensionless couplings are (almost) fixed, then anthropic arguments can be made successfully and explain why we observe a fine-tuned universe. If we start varying every single SM coupling across the Multiverse it becomes much harder to explain why we live in our universe~\cite{Hall:2014dfa, Harnik:2006vj}. Additionally, even dynamical explanations for $m_h$ that rely on a multiverse require a ``friendly landscape" that does not let everything vary from vacuum to vacuum.

This construction can be used to obtain a universe that looks like our own from theories where $\MS$ is much bigger than ${\rm meV}$ and of the weak scale. It can also be uplifted to give a solution to the doublet-triplet splitting problem via the scanning of $\mu$, as mentioned in~\cite{Arkani-Hamed:2005zuc}, and realized in an explicit dynamical model in~\cite{Csaki:2024ywk}.

It is important to keep in mind that we are not trying to describe the entire landscape. If the SM is completed into string theory there is also a much larger sector of the landscape where the dimension of spacetime, the low-energy gauge group and the particle content of the low-energy theory can change from vacuum to vacuum. The question of why the CC and Higgs mass are so small becomes much harder to address in this larger landscape. Here, until a better explanation (than Weinberg's) is found for the CC, we take the point of view that experiment is telling us that we live in a sector of the landscape where many of the properties of the SM and $\Lambda$CDM remain fixed and the larger landscape is either not populated by inflation or consists of rapidly-decaying unstable vacua.  

Another point worth keeping in mind is that we cannot draw general lessons from our field theory landscape on the distribution of string theory vacua. In the setup discussed here it is hard to scan dimensionless couplings in a $\mathcal{O}(1)$ neighborhood of their central value, and the same is true even of dimensionful ones if we do not impose {\it ad hoc} discrete symmetries. This is not necessarily the case in other realizations of the landscape. For example, in the IIB flux ensemble, $W$ does not have a Gaussian distribution, and it is easy to scan large values of the vacuum energy~\cite{Denef:2004ze, Denef:2004cf}.\rm

Even if we restrict ourselves to field theory landscapes, we should not try to draw too many general lessons from our toy model. For example one might be tempted to conclude that we always need an odd superpotential to efficiently scan the CC, but in Section~\ref{sec:joint} we construct a landscape where the CC is efficiently scanned around zero by SUSY breaking, after starting from a superpotential which is even in the fields. 

The characteristics of our toy model that we want to retain in a generic model of the landscape are the following:
\begin{enumerate}
    \item ``Friendliness" in the sense of~\cite{Arkani-Hamed:2005zuc}. We need a ``local" neighbor of the landscape that scans efficiently only dimensionful SM parameters.
    \item A large number of minima, since we need at least $\sim \MS^4/({\rm meV})^4$ to scan the CC, as explained below.
\end{enumerate}
In addition to the two points above, it is important to keep in mind that we are not going to make esplicit requirements on the range of scanning of the CC and the Higgs mass by our low energy landscapes. They are going to scan both parameters in some range, but we are not going to require that this range covers the whole distance between $M_{\rm pl}$ and the observed scale of the two parameters.

More concretely, in what follows, we keep the basic structure given by Eq.s~\eqref{eq:Wphi} and~\eqref{eq:Wmu}, i.e., we consider the leading part of the potential to be a sum of functions with two minima for $N$ decoupled scalars
\be
V \supset \sum_{i=1}^N \left|g_i H_u H_d- m^2 + 3 \lambda \phi_i^2\right|^2\, ,
\ee
this gives us a total of $2^N$ minima and scans efficiently $\Lambda$ and $m_h^2$. A few hundred scalars allow the scanning of  $\Lambda$ and $m_h^2$ from $M_{\rm Pl}$ down to their observed values. For an arbitrary cutoff $M_N$ we need $N_{\rm min}$ scalars with
\be
N_{\rm min} \simeq \log_2 \frac{M_N^6}{(m_h^2)_{\rm obs} \Lambda_{\rm obs}}\, . \label{eq:Nmin}
\ee
However, as stated above, we are never going to explicitly impose that our low energy landscapes do the whole scanning. This allows us to explore a broader class of experimental signatures, since it avoids fixing masses and VEVs of the $\phi_i$'s and also fixing $N$ to the (rather heuristic, as we are going to see just below) value in Eq.~\eqref{eq:Nmin}.

The logic behind Eq.~\eqref{eq:Nmin}, already discussed in~\cite{Arkani-Hamed:2005zuc}, is the following. $\Lambda$ and $m_h^2$ are given by a sum over at least $N$ contributions. Their distribution across the $2^N$ vacua at large $N$ is well approximated by a Gaussian, due to the central limit theorem,
\be
f(\Lambda) = \frac{2^N}{\sqrt{2\pi N} \Lambda_\sigma}e^{-\frac{\Lambda^2}{2N \Lambda_\sigma^2}}\, ,
\ee
where $\Lambda_\sigma \simeq |W_+-W_-|^2/M_{\rm Pl}^2$ in our supersymmetric example, if we set $W_{i\pm}=W_\pm$. For $\Lambda \simeq \Lambda_{\rm obs} \ll \sqrt{N} \Lambda_\sigma$ the distribution is approximately flat (we are zooming in around the maximum at $\Lambda=0$) and its cumulative distribution reads 
\be
\int d\Lambda^\prime f(\Lambda^\prime) \simeq \frac{2^{N+1}}{\sqrt{2\pi N}}\frac{\Lambda}{\Lambda_\sigma}\, .
\ee
This means that the smallest CC in this landscape is approximately
\be
\Lambda_{\rm min}\simeq \sqrt{2\pi N}\frac{\Lambda_\sigma}{2^N}\, .
\ee
Setting $\Lambda_{\rm min} = \Lambda_{\rm obs}$ and $\Lambda_\sigma = M_N^4$ we obtain Eq.~\eqref{eq:Nmin} if we neglect the factor $\sqrt{2\pi N}$. The total number of vacua in Eq.~\eqref{eq:Nmin} is the product of the number of vacua needed to get $\Lambda_{\rm min} = \Lambda_{\rm obs}$ and those needed for $(m_h^2)_{\rm min} = (m_h^2)_{\rm obs}$, since the two parameters scan independently.

The choice of decoupling the scalars from each other is made purely for simplicity, as it makes manifest the exponentially large number of minima. In general, SUSY breaking will introduce cross-couplings between the scalars\footnote{We could have introduced them also at the level of the superpotential, but we prefer to consider this simpler structure that allows to naturally keep them parametrically smaller than the rest of the $\phi_i$'s potential.}, and in studying their phenomenology, we take them into account.  At low energy, well below the SUSY breaking scale, we are going to consider the potential
\be
V_1= \sum_{i=1}^N \frac{\epsilon_i^2}{4}\left|\phi_i^2 - M_*^2\right|^2 + \sum_{i,j} \frac{\lambda_{ij}}{N} \phi_i \phi_j |H|^2 +\sum_{i,j, k,l} \frac{\lambda_{ijkl}^\prime}{N^2} \phi_i \phi_j \phi_k \phi_l + {\rm h.c.} \label{eq:V1}
\ee
The first term and the diagonal entries in $\lambda^\prime$ come from the superpotential in Eq.~\eqref{eq:Wphi}, while the diagonal $\lambda$'s come from the superpotential in Eq.~\eqref{eq:Wmu} after integrating out all heavy Higgses. The remaining terms arise after SUSY breaking. Note that they can be generated from soft terms as in Section~\ref{sec:joint}. We did not include linear and cubic terms because the large VEVs of the $\phi_i$'s will generate them, so from the point of view of understanding collider phenomenology qualitatively, we are not loosing in generality. For a study including those terms in a similar setting we refer to~\cite{Arkani-Hamed:2020yna, DAgnolo:2019cio}.

\begin{figure}
    \centering

\includegraphics[width=1.0\linewidth]{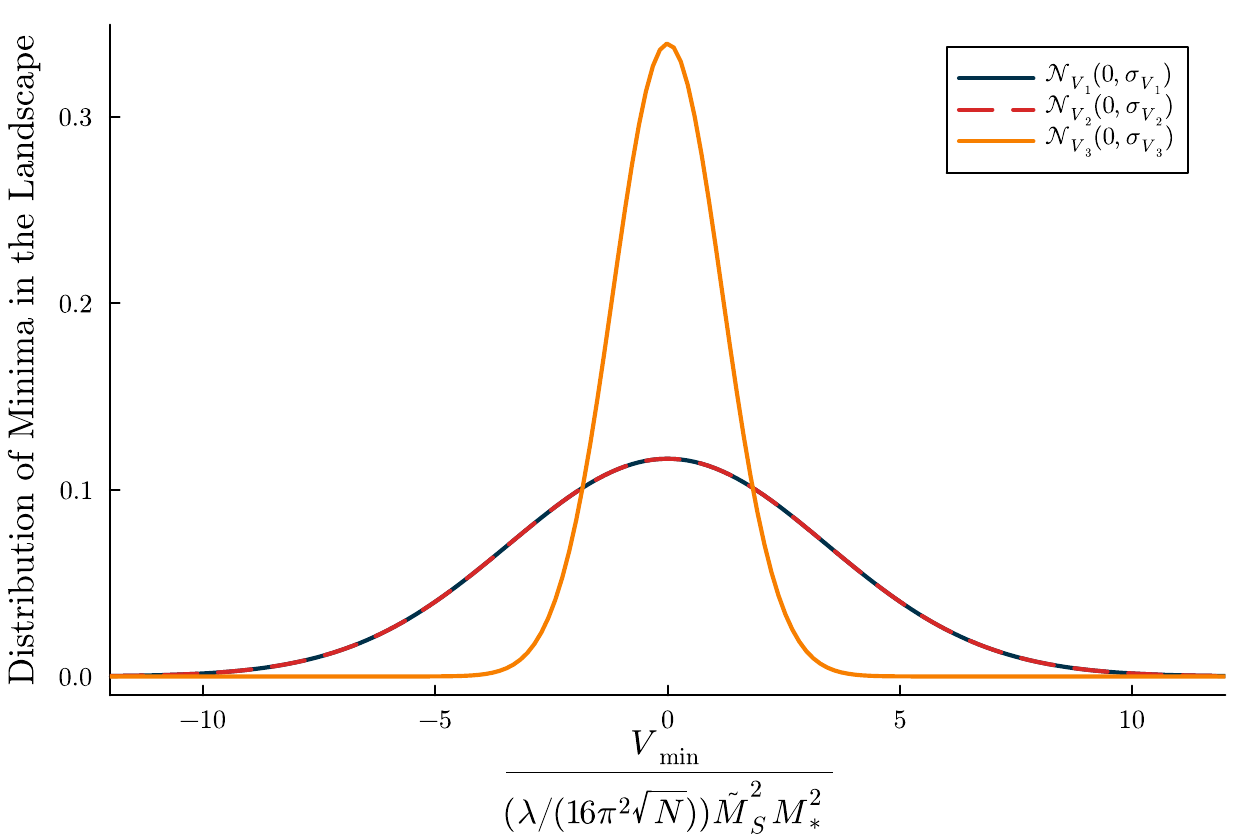}
    \caption{The vacuum energy of minima in the landscape generated by the potentials  in Eq.s~\eqref{eq:V1}, \eqref{eq:V2} and~\eqref{eq:V3}, respectively. The distribution is Gaussian, $\mathcal{N}$, with zero mean and standard deviation $\sigma_V \simeq A(\lambda/(16\pi^2\sqrt{N})) \MS^2 M_*^2$, where $A$ is an $\mathcal{O}(1)$ number that depends on the potential. We comment on the typical value of $\langle |H|^2 \rangle$ (which is not simply the Higgs VEV squared), and its UV-sensitivity around Eq.~\eqref{eq:muprime}.}
    \label{fig:Vd}
\end{figure}

It is important to notice that Eq.~\eqref{eq:V1} and our toy models of the landscape introduced in the following do not scan the CC as in Eq.~\eqref{eq:SUSYCC}. We imagine that $|W|^2/M_{\rm Pl}^2 \ll V_{\rm min}$, so the vacuum energy is dominated by SUSY breaking and is of order $V_{\rm min}\simeq (\lambda/(16\pi^2\sqrt{N})) \MS^2  M_*^2$.  This is shown in Fig.~\ref{fig:Vd}, where it is clear that our toy model scans the CC symmetrically around zero.

Going back to the details of Eq.~\eqref{eq:V1}, to avoid complications from the cross-couplings, we are going to assume that they do not affect most of the minima
\be
\epsilon^2 \gg \lambda \frac{v^2}{M_*^2}, \quad \epsilon^2 \gg \lambda^\prime\, .
\ee
To ensure that the potential is technically natural (i.e. there are no large loop corrections to the $\phi_i$'s masses and quartic couplings from the cross-couplings or the couplings to the Higgs) we also need
\be
\epsilon^2 \gg \frac{\lambda}{16\pi^2 N} \frac{\MS^2}{M_*^2} \, , \quad \lambda^\prime \gg \frac{\lambda^2}{16\pi^2}\log\frac{\MS^2}{m_h^2}\, ,
\ee
assuming a cutoff for Higgs loops at $\MS$.
The potential in Eq.~\eqref{eq:V1} is our starting point for a phenomenological study of a generic ``friendly" landscape. Its most important phenomenological feature is the mass distribution that it generates. For $\epsilon_i$ changing at $\mathcal{O}(1)$ between different scalars, we are going to have physical masses distributed in the interval $[a,b]\times \epsilon M_*$, where $a$ and $b$ are $\mathcal{O}(1)$ numbers.

It is also important to consider the case where the sign of $m^2$ in the superpotential changes from scalar to scalar. Imagine that $N/2$ scalars have positive mass squared parameters and the other half negative mass squared parameters. More precisely, we are going to imagine that the $m^2$ coefficients are distributed uniformly between $-\epsilon^2 M_*^2$ and $\epsilon^2 M_*^2$. The total number of minima will be exponentially smaller ($2^{N/2}$ instead of $2^N$), but now the physical masses of the scalars in $V_1$ span the interval $[c/\sqrt{N}, d]\times \epsilon M_*$, where $c, d$ are again $\mathcal{O}(1)$, leading to a rather different collider phenomenology, as we will see in Section~\ref{sec:pheno}. The appearance of $1/\sqrt{N}$ in the mass spectrum is just a consequence of having assumed $N$ states uniformly distributed within the mass squared interval. 

The phenomenological differences are sufficiently important that we write a new potential
\be
V_2= \sum_{i=1}^N \frac{\epsilon_i^2}{4}\left|\phi_i^2 - \delta _i M_*^2\right|^2 + \sum_{i,j} \frac{\lambda_{ij}}{N} \phi_i \phi_j |H|^2 +\sum_{i,j, k,l} \frac{\lambda_{ijkl}^\prime}{N^2} \phi_i \phi_j \phi_k \phi_l + {\rm h.c.} \label{eq:V2}\, ,
\ee
that differs from $V_1$ only by the sign of $\delta_i=\pm 1$ and the range of the $\epsilon_i$'s that can span the interval $[1/\sqrt{N},1]\times \epsilon$. This potential is scanning the CC symmetrically around zero as shown in Fig.~\ref{fig:Vd}. 

The last generic model that we would like to mention is one where there is some notion of locality in theory space, for instance, because of an extra dimension~\cite{Arkani-Hamed:1999ylh}. In this model, each scalar couples only to its two nearest neighbors, so that
\be
\lambda_{ij}&=&\lambda(a_1\delta_{i,j+1}+a_2\delta_{i,j-1}+a_3\delta_{i,j})\, , \label{eq:V3}  \\
\lambda_{ijkl}^\prime&=&\lambda^\prime(b_1\delta_{i,j+1}+b_2\delta_{i,j-1}+b_3\delta_{i,j})(b_4\delta_{i,k+1}+b_5\delta_{i,k-1}+b_6\delta_{i,k})(b_7\delta_{i,l+1}+b_8\delta_{i,l-1}+b_9\delta_{i,l})\, , \nn
\ee
where the $a_i$'s and $b_i$'s are $\mathcal{O}(1)$ numbers. We will not discuss the phenomenology of this model in detail in what follows because most generic mass spectra lead to simple two-body decays for most of the scalars, so the phenomenology of this model is not dissimilar to that of $N$ copies of a single scalar mixing with the Higgs, whose phenomenology has already been studied in detail (see for instance~\cite{Ferber:2023iso} and its references). This concludes the list of our phenomenological toy models of a ``local" neighbor of the landscape. The potential from Eq.~\eqref{eq:V3} gives a narrower distribution for the CC in the landscape compared to Eq.s~\eqref{eq:V1} and~\eqref{eq:V2}, but it is still scanning the CC symmetrically around zero as shown in Fig.~\ref{fig:Vd}.

So far, we have assumed by fiat that the $\phi_i$'s are around the weak scale. However, there is a simple symmetry reason that could explain why they are at the weak scale or below. Imagine that the new scalars are charged under a discrete subgroup of the $U(1)$ PQ symmetry of the SM Higgs sector. The symmetry might forbid the $m^2 \phi$ term in Eq.~\eqref{eq:Wphi} while leaving $\phi^3$ invariant. If the leading term that scans the $\mu$ parameter 
\be
W \supset \sum_{i=1}^N g_i \phi_i H_u H_d\, ,
\ee
is also invariant, the masses of the new scalars are going to be $\mathcal{O}(g_i v^2)$ in our universe. In this case, we would need to invoke another sector of the landscape to scan  $\Lambda$ and $m_h^2$ down to the weak scale, but the scalars in this low-energy sector of the landscape are naturally below the weak scale, since in a perturbative theory we expect $g_i \sim 1/\sqrt{N}$. 

Before discussing the phenomenology of our landscape models, we describe a special limit of $V_1$ that leads to a joint solution of the CC and electroweak hierarchy problems. We leverage a combination of high-energy supersymmetry and Weinberg's anthropic argument for the CC.

\section{A Joint Solution to the Cosmological Constant and Electroweak Hierarchy Problems}\label{sec:joint}
In this Section, we discuss a model that can explain at the same time the value of the CC and that of the Higgs mass using only Weinberg's anthropic argument for the CC~\cite{Weinberg:1987dv}.

We start by outlining the basic logic of this idea, which is similar, but not identical, to~\cite{Arkani-Hamed:2020yna}. In our model, SUSY solves both fine-tuning problems down to a scale $\MS \gg m_h^{\rm exp}$ where SUSY breaking takes place. Below $\MS$, the CC and the Higgs mass are scanned\footnote{Note that the typical value of the CC in this theory is $\sim 16 \pi^2\MS^4$ while the Higgs mass is one loop below at $\sim \MS^2$, with its detailed value depending on the SUSY-breaking mediation mechanism. This is because we take $\MS$ to be the typical mass of a SM superpartner and we do not know how to break supersymmetry with tree-level couplings to the MSSM in an experimentally viable way, see~\cite{Martin:1997ns} for a review.} by a low energy landscape. This landscape has a special structure: only universes with a small and negative $m_h^2$ have enough non-degenerate vacua to scan also the CC. In all other universes,  we either loose some minima in the landscape, because of a large tadpole induced by $m_h^2$, or these minima become degenerate.

To be more concrete, we can write the superpotential 
\be
W = \mu H_u H_d + \frac{\epsilon^2}{2}\sum_{i=1}^N X_i(\phi_i^2- M_*^2) \, , 
\label{eq:W}
\ee
where the $N$ new scalars $\phi_i$ realize our low-energy landscape and the $X_i$'s are a set of auxiliary fields. We take the VEV $M_*$ and the coupling $\epsilon^2$ to be equal for all of the scalars just for simplicity, but more general choices do not affect our mechanism.

We are going to see in the following that this landscape can explain the value of the Higgs mass and of the CC. 
We take the fields in the landscape to be parametrically lighter than the SM Higgs. This is required to explain the CC, as we discuss in the following.  As we explained in the previous section,  the choice $m_\phi < m_h$ is technically natural.

In the absence of SUSY breaking the SM is coupled to the low energy landscape only through gravity. We introduce a coupling between the two sectors through SUSY breaking. We imagine that SUSY breaking respects the $Z_4$ symmetry
\be
Z_4 \; : \; H_u H_d \to - H_u H_d\, , \quad \phi_i \phi_j \to - \phi_i \phi_j\, , \quad X_i \to - X_i\, ,\label{eq:Z4}
\ee
except for a soft breaking coming from the usual $B\mu$ term in the Higgs potential.
For simplicity, we also assume that the same spurion that breaks the $\phi_i$'s shift symmetry in the superpotential also governs the SUSY-breaking corrections to the $\phi_i$'s potential. With these choices, the only soft SUSY breaking terms that we can write down to connect the two sectors are
\be
V_{\phi H} = \frac{M_X^2}{2} \sum_i X_i^2 + A_H X H_u H_d + \frac{A_\phi}{N} X \sum_{i,j} a_{ij}\phi_i \phi_j + {\rm h.c.}\, , \label{eq:VphiH}
\ee
where $A_\phi \propto \epsilon^2$, $X\equiv (1/\sqrt{N}) \sum_i X_i$ and $A=\{a_{ij}\}_{i,j}$ is a real, symmetric and non-singular matrix of $\mathcal{O}(1)$ numbers. For simplicity we take the $X_i$'s masses and couplings to be all equal. In the following, we are going to take $M_X \gg \epsilon M_*$ so that we can integrate out the $X_i$'s and simplify our discussion. This is technically natural because of the approximate shift symmetry on the $\phi_i$'s. 

Additionally we imagine that $A_\phi \ll A_H$. This is technically natural because $A_\phi$ breaks a second $Z_4$ symmetry, that we call $Z_4^H$, acting as
\be
Z_4^H \; : \; H_u H_d \to - H_u H_d\, \quad X_i \to - X_i\, ,
\ee
which is preserved by $A_H$. In this limit, the VEV of $X$ is mainly induced by EW symmetry breaking
\be
\langle X \rangle \simeq -A_H \frac{\langle H_u H_d\rangle}{M_X^2}\, . \label{eq:Xvev} 
\ee
We are going to see in a moment why this is important. For simplicity, we take $\langle X \rangle \ll M_*$, so the dominant contribution to the $\phi_i$ masses comes from the superpotential. To complete the list of SUSY-breaking terms in the theory we add the usual soft terms in the Higgs potential
\be
V_{H_u H_d} = m_{H_u}^2 |H_u|^2 + m_{H_d}^2 |H_d|^2 + (B\mu H_u H_d+{\rm h.c.})\, .
\ee
Since we have introduced in the theory a $B\mu$ term for the Higgses, we could also include a soft breaking of the $Z_4$ in Eq.~\eqref{eq:Z4} proportional to $\phi_i \phi_j$, but we are interested in the limit where $B\mu$ is small enough as not to affect the $\phi_i$'s potential, so it is technically natural to omit this term altogether. In the next Section we discuss the upper bound on $B\mu$ and its origin.

If we integrate out the $X_i$'s and $H_{u,d}$ we obtain the low-energy scalar potential
\be
V= \sum_i \frac{\epsilon^2}{4}\left|\phi_i^2 - M_*^2\right|^2 +\frac{\lambda}{N} \mu^2_H\sum_{i,j}a_{ij}\phi_i \phi_j+  \frac{\lambda^{\prime}}{N^2}\left(\sum_{i,j}a_{ij}\phi_i \phi_j\right)^2 + ... \; ,\label{eq:V}
\ee
where $\mu_H^2$ is a quantity that can be sensitive to the weak scale, defined as
\be
\mu_H^2 \equiv \langle H_u H_d \rangle\, . \label{eq:muH}
\ee
We discuss $\mu_H$ in detail in the next Section. The two quartics $\lambda, \lambda^\prime$ come from the interactions between the two sectors mediated by the $X$ boson, $\lambda \propto A_H A_\phi/M_X^2$, $\lambda^\prime \propto A_\phi^2/M_X^2$. We see from Eq.~\eqref{eq:V} that the $X$ VEV in Eq.~\eqref{eq:Xvev} makes the low-energy scalar potential sensitive to the weak scale through $\mu_H^2$. This is the main ingredient of our mechanism. 

\begin{figure}
    \centering    \includegraphics[width=1.0\linewidth]{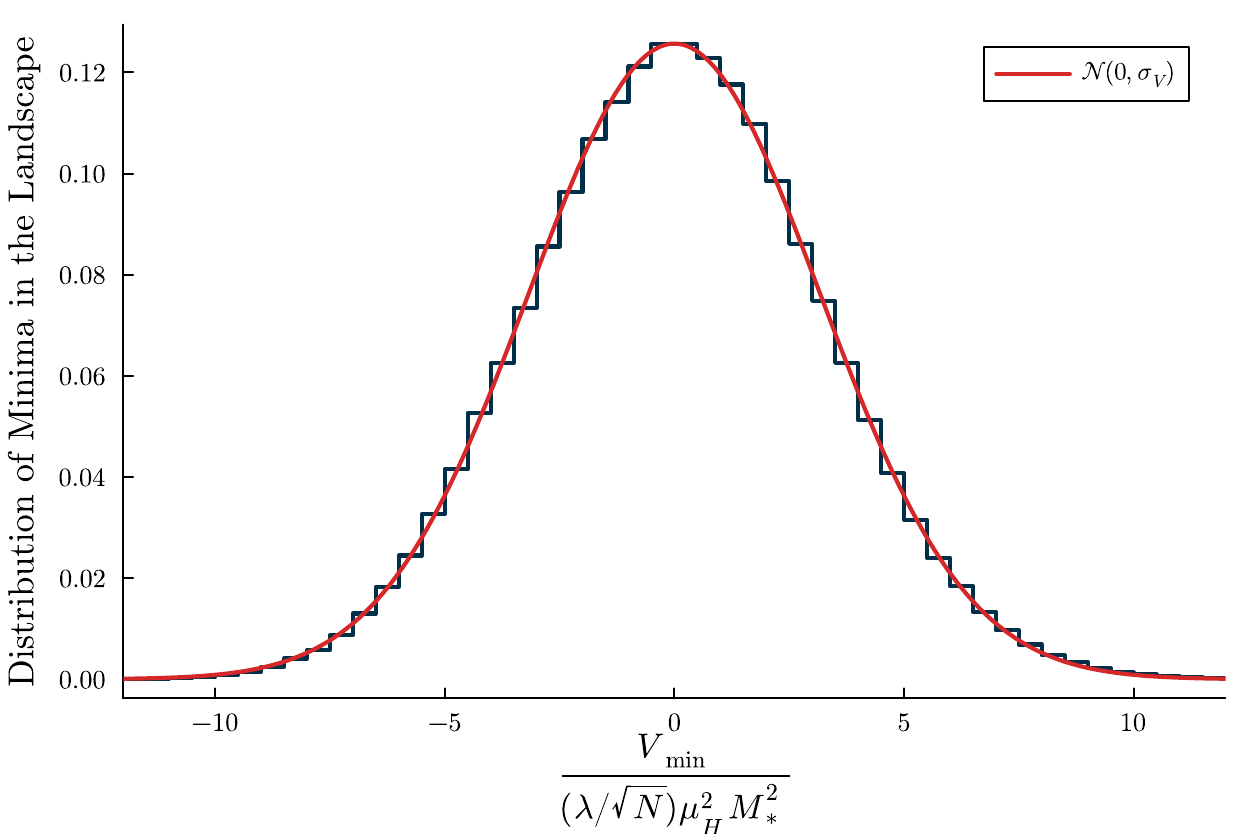}
    \caption{The vacuum energy of minima in the landscape generated by the potential in Eq.~\eqref{eq:V} follows a normal distribution with zero mean and standard deviation $\sigma_V \simeq (\lambda/\sqrt{N})\mu_H^2 M_*^2$, where $M_*$ is the cutoff of the theory and $\mu_H^2$ is defined in Eq.~\eqref{eq:muH}.}
    \label{fig:Vjoint}
\end{figure}

 In the following we show that this potential implements our selection mechanism for the CC and the Higgs mass. We break down the discussion into two steps. First, we discuss the dependence of Eq.~\eqref{eq:V} on the weak scale, then we show the connection between $m_h^2$ and the CC. At this stage we can already note that Eq.~\eqref{eq:V} is a special case of Eq.~\eqref{eq:V1} and also scans the CC symmetrically around zero, as shown in Fig.~\ref{fig:Vjoint}.
 
\begin{figure}[!t]
    \centering
    \includegraphics[width=1.0\linewidth]{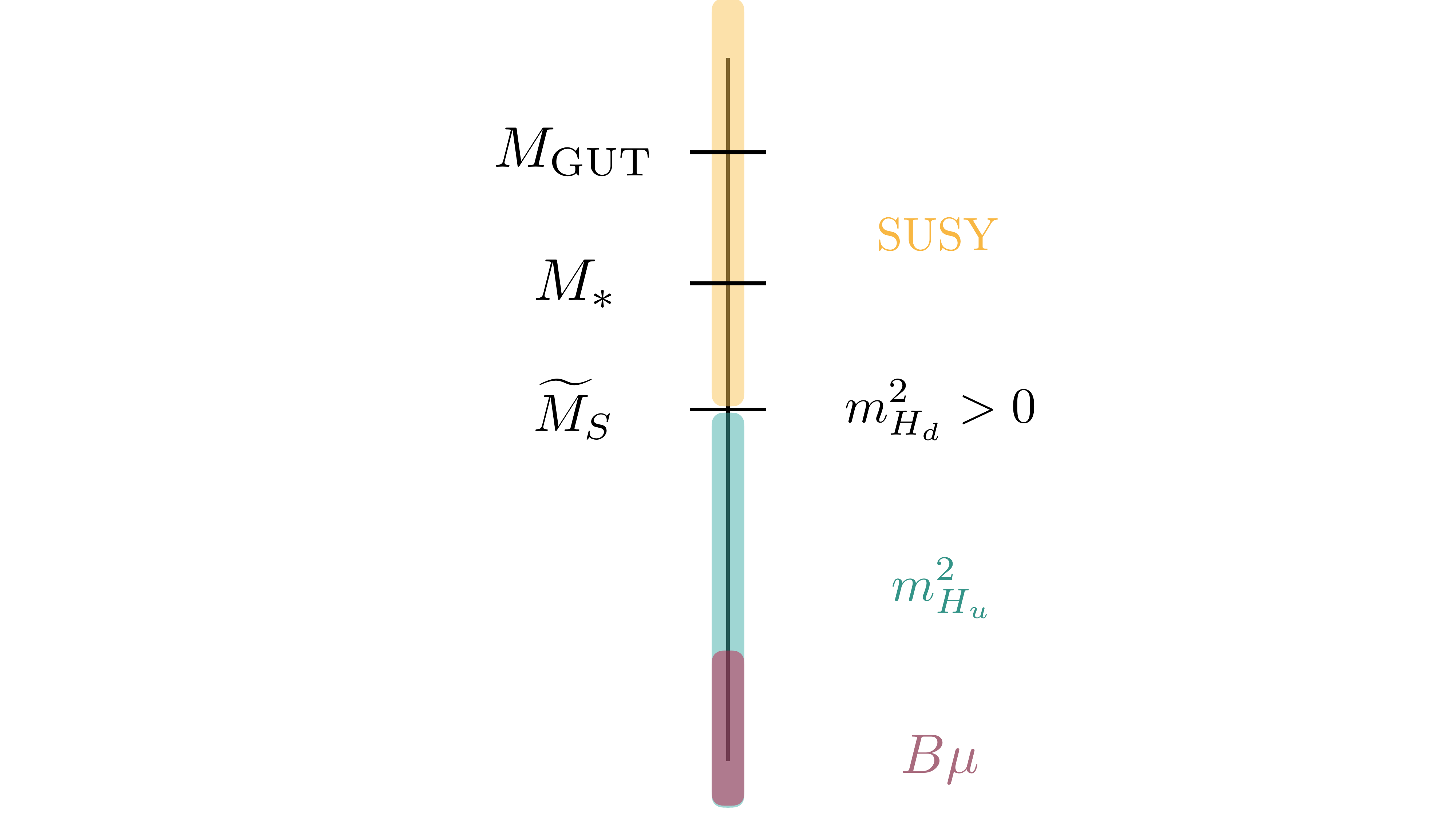}
    \caption{Schematic representation of the energy scales in the Higgs sector of the model in Section~\ref{sec:joint} that can jointly solve the CC problem and the electroweak hierarchy problem.}
    \label{fig:sketch1}
\end{figure}

\subsection{Sensitivity to the Weak Scale}\label{sec:weak}
The crucial parameter in Eq.~\eqref{eq:V} is $\mu^2_H$ which is given by
\be
\mu^2_H = \langle H_u \rangle \langle H_d \rangle + c\frac{B\mu}{16\pi^2}\log \frac{\MS^2}{m_H^2}\, ,
\ee
where $m_H$ is a combination of physical Higgs masses and $c$ is a $\mathcal{O}(1)$ coefficient.
If we take the $B\mu$ term in $V_{H_u H_d}$ to be sufficiently small, $\mu_H$ becomes a function of the $H_{u,d}$ VEVs. This requires 
\be
B\mu \lesssim 16\pi^2\frac{\langle H_u \rangle \langle H_d \rangle}{ \log{\frac{\MS^2}{|m_H|^2}}} = 16\pi^2\frac{v^2}{ \log{\frac{\MS^2}{m_H^2}}}\frac{\tan{\beta}}{\tan^2{\beta}+1} \lesssim   \frac{(500\; \textrm{GeV})^2}{\log\frac{\MS^2}{(10\,{\rm TeV})^2}} \; . \label{eq:Bmu}
\ee
The second inequality ensures that in universes like our own or with a larger EW-breaking scale, the $B\mu$ contribution to the VEV is subleading.   In this limit, $\mu_H^2 \simeq \langle H_u \rangle \langle H_d \rangle$ is sensitive to the weak scale. In universes without EW symmetry breaking or with smaller VEVs than our own Eq.~\eqref{eq:Bmu} ensures that $B\mu$ cannot split the $\phi_i$'s minima enough to scan the CC, as we discuss in the next Section. 

In what follows we consider the standard decoupling limit of the MSSM 2HDM where $m_{H_d}^2 \simeq \MS^2$ is large and positive, $B\mu$ is small and $m_{H_u}^2$ is scanned in the Multiverse to reproduce the weak scale in some vacua. We take the scanning of $B\mu$ to be at most comparable to the upper bound in Eq.~\eqref{eq:Bmu}. As usual, it is fine to take $B\mu$ smaller than other SUSY breaking terms in the Higgs sector because it breaks a PQ symmetry of which $H_u H_d \to - H_u H_d$ is a subgroup. This structure is depicted schematically in Fig.~\ref{fig:sketch1}.

To conclude, note that to retain the sensitivity to the weak scale,  we do not want the $B\mu$ term generated by $\lambda$ to exceed the upper bound in Eq.~\eqref{eq:Bmu}. This implies
\be
\lambda \lesssim \frac{16 \pi^2}{c_1}\frac{B\mu^{\rm max}}{\MS^2} \simeq \frac{0.16}{c_1}\left(\frac{10\;{\rm TeV}}{\MS}\right)^2\, .
\ee
We do not compute $c_1$ because a much stronger upper bound on $\lambda$ arises from a different requirement on the model discussed in the next Section.

Additionally, we are imagining that a different sector of the landscape is scanning the Higgs potential to tune down one Higgs below $\MS$. The low energy sector that scans the CC only couples to $H_u H_d$ and scans $B\mu$ by less than its upper bound in Eq.~\eqref{eq:Bmu}. Relaxing this condition introduces terms such as $|H|^2 \phi_i \phi_j$ that could make the low-energy landscape insensitive to the Higgs VEV. In particular, they lead to
\be
V\supset (\mu'_H)^2 \phi_i \phi_j, \quad (\mu'_H)^2 = \frac{\MS^2}{16\pi^2}+\langle H \rangle^2, \label{eq:muprime}
\ee
where the first term in $(\mu'_H)^{2}$ comes from closing the Higgs loop with the $|H|^2 \phi_i  \phi_j$ coupling, 
and the second term comes from the tree-level contribution of the Higgs VEVs.  Hence, the potential would be insensitive to the electroweak symmetry breaking if the SUSY breaking scale is too high. For a more detailed discussion on ``trigger" operators, i.e., operators whose VEV is sensitive to EW symmetry breaking we refer to~\cite{Arkani-Hamed:2020yna, TitoDAgnolo:2021pjo, Catinari:2025zya}.

\subsection{Joint Selection of $m_h^2$ and the CC}

In order to develop some analytic intuition on the mechanism, we start by analyzing a simplified potential with $\lambda^\prime=0$. In this limit Eq.~\eqref{eq:V} reads
\be
    V_{s}&=& V_0 + V_H\, , \nn \\ V_0&=& \sum_i \frac{\epsilon^2}{4}\left(\phi_i^2 - M_*^2\right)^2\, , \quad
    V_H=\frac{\lambda}{N} \mu^2_H\sum_{i,j}a_{ij}\phi_i \phi_j \; .
    \label{eq:Vsimplified}
\ee
Understanding the mechanism is a simple exercise in determining the vacua of $V_s$. It is useful to first consider $V_0$ and $V_H$ separately.

$V_0$ has $2^N$ minima at $\phi_i = \pm M_*$ that are all degenerate, $V_0^{\rm min} = 0$, and one maximum at $\phi_i =0$, with potential $V_0^{\rm max} =N\epsilon^2 M_*^4/4$. When $V_0$ dominates,  we have enough minima to scan the CC (since we assume $2^{-N} \MS^4 \lesssim {\rm meV}^4$), but they are all degenerate. 
Now consider $V_H$. Since we took the matrix of the $a_{ij}$ coefficients to be non-singular (i.e. $\det A \neq 0$), if $V_H(|\phi_i| \lesssim M_*) \gg V_0(|\phi_i| \lesssim M_*)$ we have a single stationary point at $\phi_i =0$. So we want to be in the regime where $V_0$ dominates in the region $|\phi_i| \lesssim M_*$. This can be translated into two quantitative conditions on $\mu_H^2$. If we do not want to loose {\it any} of the $2^N$ minima we need
\be
\mu_H^2 \ll \mu_{\rm max}^2 \simeq  \frac{\epsilon^2 M_*^2}{4 \lambda}\; . \label{eq:muMax}
\ee
For most minima the sum over $a_{ij} \phi_i \phi_j$ scales as $N$ rather than $N^2$ (i.e. the signs of the $\phi_i$'s VEVs alternate) and we have the looser condition $\mu_H^2 \ll N \mu_{\rm max}^2$, but in what follows we are going consider the stronger condition in Eq.~\eqref{eq:muMax}.

At the same time we want $V_H$ to split the minima enough to scan the CC down from $\MS^4$ to its observed value. This requires 
\be
\mu_H^2 \gtrsim \mu_{\rm min}^2 \simeq \frac{(16 \pi^2)^2 \MS^4}{\lambda M_*^2}\, . \label{eq:muMin}
\ee
This condition descends from the splitting between the minima induced by $V_H$. For $\mathcal{O}(1)$ coefficients $a_{ij}$, the minima have a Gaussian distribution with zero mean and standard deviation $(\lambda/\sqrt{N})\mu_H^2 M_*^2$. Eq.~\eqref{eq:muMin} is just the requirement that the maximal CC be smaller than the standard deviation of the distribution of vacuum energies shown in Fig.~\ref{fig:Vjoint}. To have at least one minimum with the observed CC, we also need enough minima. In practice we need both Eq.~\eqref{eq:muMin} and enough minima to scan the observed CC, $\Lambda_{\rm obs}$, i.e. at least
\be
N \simeq \log_2 \frac{(16 \pi^2)^2 \MS^4}{\Lambda_{\rm obs}}\, , \quad \Lambda_{\rm obs}\simeq {\rm meV}^4\, , \label{eq:enoughN}
\ee
new scalars, where for simplicity we have approximated the distribution of vacua around the maximum of the Gaussian in Fig.~\ref{fig:Vjoint} with a uniform distribution. Notice that compared to the initial discussion in Section~\ref{sec:generic} we are assuming $|W|^2/M_{\rm Pl}^2 \ll \MS^4$, which explains our qualitatively different conclusions. 

Together Eq.s~\eqref{eq:muMax}, \eqref{eq:muMin} and \eqref{eq:enoughN} select the weak scale. If we want the CC to have its observed value in at least one universe, Eq.s~\eqref{eq:muMax}, \eqref{eq:muMin}, and the discussion in Section~\ref{sec:weak} imply
\be
\mu_{\rm min}^2\lesssim \langle H_u \rangle \langle H_d \rangle \ll \mu_{\rm max}^2\, .
\ee
Given our assumptions on the Higgs sector in Secion~\ref{sec:weak} we can rewrite the previous equation as
\be
\tilde \mu_{\rm min}^2\lesssim \langle H_u \rangle^2 +\langle H_d\rangle^2 \ll \tilde \mu_{\rm max}^2\, ,
\ee
where
\be
\tilde \mu_{\rm min, max} \equiv \mu_{\rm min, max} \frac{B\mu^2 +m_{H_d}^4}{B\mu \, m_{H_d}^2}\, ,
\ee
so our limits on $\mu_H^2$ are directly bounding the EW symmetry breaking VEV from above and below, provided that the ratio $B\mu/m_{H_d}^2$ does not scan appreciably throughout the Multiverse. 
Eq.s~\eqref{eq:muMax} and~\eqref{eq:muMin} have several phenomenological consequences. An interesting one is the relation between the minimal mass of the new scalars and the SUSY breaking scale. At the minima of the potential $V_s$, the term $V_0$ dominates so $m_\phi \simeq \epsilon M_*$. Putting Eq.s~\eqref{eq:muMax} and~\eqref{eq:muMin} together we obtain 
\be
m_\phi \simeq \epsilon M_* \gg \frac{32 \pi^2 \MS^2}{M_*}\, . \label{eq:mass_bound}
\ee
This lower bound does not look particularly instructive, since we can take $M_*$ as large as we want, but it has consequences for the maximal allowed size of $\lambda^\prime$ that in turn determines the length of the decay chains of the $\phi_i$'s.

To complete this discussion we can find the upper bound $\lambda^\prime$ that allows to preserve our selection mechanism. We want $V_H$ to be the main source of splitting between the $V_0$ minima; otherwise we would not have any lower bound on $\langle H_u \rangle$, and consequently, no upper bound on large positive Higgs mass squared. At tree-level, this means
\be
\lambda^\prime \lesssim \lambda \frac{\mu_H^2}{M_*^2} \ll \lambda \frac{\mu_H^2 m_\phi^2}{(16 \pi^2)^2 \MS^4} \lesssim 10^{-12} \lambda \left(\frac{10\;{\rm TeV}}{\MS}\right)^4 \, , \label{eq:lambda_bound}
\ee
where we used Eq.~\eqref{eq:mass_bound} together with Eq.s~\eqref{eq:muMax} and~\eqref{eq:muMin} to obtain the second inequality, and additionally $m_\phi \lesssim m_h$. 
We also want to avoid the loop correction to $V_H$ exceeds its tree-level value. However, this gives a subleading bound to Eq.~\eqref{eq:lambda_bound} 
\be
\lambda^\prime \lesssim 16 \pi^2 \frac{\lambda}{N} \frac{\mu_H^2}{\MS^2} \simeq 10^{-4} \lambda \left(\frac{10\;{\rm TeV}}{\MS}\right)^2\, .
\ee
Additionally, since $\lambda^\prime$ is generated at one loop from $\lambda$ we also have the naturalness constraint 
\begin{equation}
    \lambda^\prime \gtrsim \frac{\lambda^2}{16\pi^2}\log \left(\frac{\MS^2}{|m_H|^2} \right) \;.
\end{equation}
Together with \eqref{eq:lambda_bound} this leads to 

\begin{equation}
    \lambda \lesssim 10^{-11}\left(\frac{10\;{\rm TeV}}{\MS}\right)^4 \frac{1}{\log \left(\frac{\MS}{10\;{\rm TeV}}\right)^2}  \, .
\end{equation}
This is not necessarily a problem as we do not have an upper bound on $M_*$ and the effective coupling $\theta$ (defined in Eq.~\eqref{mixingangle}) between the SM and the landscape can be large. However Eq.~\eqref{eq:lambda_bound} makes two body decays to the SM the preferred decay channel, as we are going to see in the next Section. Therefore current low-multiplicity searches are already sensitive to this model and there is not much to gain in designing more sophisticated analyses for high-multiplicity final states.

\section{Phenomenology}\label{sec:pheno}

\begin{figure}[!t]
    \centering   \includegraphics[width=1.0\linewidth]{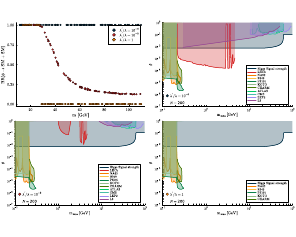}
    \caption{The upper left panel shows an example of a mass distribution and the branching ratios of the scalars into a two body SM final state. The three different colors show different choices of the $\lambda^\prime/\lambda$ ratio. The remaining panels show existing experimental constraints on the heaviest scalar in the spectrum for the same three choices of $\lambda^\prime/\lambda$. The constraints get progressively weaker as $\lambda^\prime$ increases and long cascade decays start to dominate the branching ratios. The lowest mass region, up to $0.2$ GeV is constrained by fixed target experiments~\cite{CHARM:1985anb, BNL-E949:2009dza, Gorbunov:2021ccu, NA62:2020pwi, KOTO:2020prk, NA62:2021zjw, MicroBooNE:2021sov, MicroBooNE:2022ctm}, above that threshold LHCb searches for $B\to \phi K$ dominate the bounds~\cite{LHCb:2015nkv, LHCb:2016awg, Winkler:2018qyg}. At higher masses, we show direct ATLAS, CMS and LEP searches~\cite{L3:1996ome, LEPWorkingGroupforHiggsbosonsearches:2003ing, CMS:2018zvv, ATLAS:2021hbr} and bounds on the Higgs signal strength in dark blue.}  
    \label{fig:money}
\end{figure}

The main message of this Section is summarized in Fig.~\ref{fig:money}, where we show current constraints on the heaviest scalar in the spectrum for three choices of model parameters. If we turn off the cross couplings of the scalars, $\lambda^\prime/\lambda \lesssim 10^{-6}$, low energy landscapes are already probed by current collider searches, as shown in the upper right panel of the Figure. As we increase the cross couplings, collider constraints get progressively weaker and disappear completely when $\lambda^\prime/\lambda \simeq 1$ (lower right panel of the Figure). This is not at all surprising, since large $\lambda^\prime/\lambda$ corresponds to high-multiplicity final states, due to the cascade decays of the scalars within their sector. These final states contain a large number of particles, each with a modest amount of energy. Current searches do not capture this cascading regime, which in our models is quite generic. $\lambda^\prime$ is generated at one loop by $\lambda$ while the converse is not true, and parametrically it is more natural for the new scalars to cascade within their sector, when this is permitted by phase space. Note that the results in this Section apply to all the models described in Sections~\ref{sec:generic} and~\ref{sec:joint}, with the exception of the nearest neighbors couplings in Eq.~\eqref{eq:V3}. We do not consider that model because it generically leads to simple two-body final states whose phenomenology has already been studied in detail (see, for instance~\cite{Ferber:2023iso}).

In the following we elaborate more on these points and discuss in detail the existing constraints on landscapes and the phenomenology of their cascade decays. Note that we choose to focus on what we think is the most challenging mass range for these new landscapes. We consider scenarios where $m_\phi \lesssim 200$~GeV so that the total energy from the decays of the new scalars is not enough by itself to decisively reduce the SM backgrounds. At the lower end of the spectrum, we stop at $m_\phi \gtrsim 0.1$~GeV, since around these values inclusive searches at beam-dump experiments are already sensitive to our low energy landscapes.

\subsection{Existing Experimental Constraints}
In Fig.~\ref{fig:money}, we show existing constraints applied to the heaviest scalar in our spectrum. We choose the heaviest scalar because it illustrates in a particularly simple way the importance of high multiplicity final states for the detection of a landscape. We display the constraints in terms of the mixing angle $\theta$ between the SM and a single scalar in the landscape, whose typical value is
\begin{equation}
      \theta_i \simeq \theta \simeq \frac{2\lambda M_* v}{\sqrt{N} |m_h^2 - m_\phi^2|} \; ,
      \label{mixingangle}
\end{equation}
where we approximated the sum $(1/N)\sum_j \lambda_{ij} \langle \phi_j \rangle \simeq \lambda M_*/\sqrt{N}$. The logic is that most minima have alternating signs for the $\phi_i$'s VEVs, i.e. the probability distribution of the sum $\sum_j \langle \phi_j \rangle$ has zero mean and standard deviation of order $\sqrt{N} M_*$, so in a typical minimum of the landscape we expect $\sum_j \langle \phi_j \rangle \simeq \sqrt{N} M_*$.

Scalar masses above GeV are mainly constrained by collider searches at the LHC and LEP. From LEP we show the two analyses~\cite{L3:1996ome, LEPWorkingGroupforHiggsbosonsearches:2003ing} that searched for new scalars produced by Higgstrahlung. From the LHC, we show in dark blue the constraints from Higgs coupling measurements and in light blue (marked as ATLAS and CMS in the Figure) direct collider constraints from~\cite{L3:1996ome, LEPWorkingGroupforHiggsbosonsearches:2003ing, CMS:2018zvv, ATLAS:2021hbr} that target four body final states (mainly $\tau \tau \mu\mu$ and $bb\mu\mu$).  The strongest indirect constraint comes from the ATLAS and CMS combined measurement of the overall signal strength of the Higgs~\cite{ATLAS:2022vkf, CMS:2022dwd},
\begin{equation}
    \mu = 1.03 \pm 0.06\, .
\end{equation}
The new scalars change $\mu$ by reducing the Higgs width to the SM as
\begin{align}
\Delta\mu &\simeq
\frac{\Gamma_h^{\mathrm{SM}} (1-4 N \theta^2)}
{\Gamma_h^{\mathrm{SM}} (1-2 N \theta^2) + \sum_{i,j} \Gamma_{h\rightarrow \phi_i \phi_j}} , \notag \\[1ex]
\Gamma_{h\to \phi_i \phi_j} &\simeq
\frac{\bigl(2 \lambda_{ij} (v+\sqrt{N}\,\theta M_*)\bigr)^2}{16\pi m_h N^2}
\sqrt{\left(1-\frac{(m_{\phi_i}+m_{\phi_j})^2}{m_h^2}\right)
\left(1-\frac{(m_{\phi_i}-m_{\phi_j})^2}{m_h^2}\right)}\, ,
\label{eq:Hphiphi}
\end{align}
where we show the result at leading order in the small $\theta$ limit (note that $\sqrt{N} \theta \to$~const. in the large-$N$ limit). We plot the constraint from the Higgs signal strength in dark blue in all Figures.
 In red we display constraints from $B\to K \phi$ decays in LHCb~\cite{LHCb:2015nkv, LHCb:2016awg, Winkler:2018qyg}, and in green, yellow and orange different fixed target experiments~\cite{CHARM:1985anb, BNL-E949:2009dza, Gorbunov:2021ccu, NA62:2020pwi, KOTO:2020prk, NA62:2021zjw, MicroBooNE:2021sov, MicroBooNE:2022ctm} looking for displaced two-body decays or missing energy.

Fig.~\ref{fig:money} shows that fixed target experiments and indirect constraints from the Higgs signal strength retain most of their sensitivity when the scalars decay through long cascades, as displayed in the two lower panels of the Figure. Direct searches at ATLAS, CMS, LHCb and at LEP are instead mostly sensitive to low-multiplicity final states and are effective only when the cross coupling between the scalars is small $\lambda^\prime/\lambda \lesssim 10^{-6}$ (upper right corner of the Figure). However, the upper left panel of Fig.~\ref{fig:money} shows that existing collider searches are always sensitive to the lightest scalars in the spectrum, regardless of the value of $\lambda^\prime/\lambda$. This happens because of phase space, since some number of the lightest scalars can only decay directly to the SM. In deriving these limits from the direct searches, we consider that they are only sensitive to the decay channel which they are optimized for. In practice, a more detailed analysis could reveal that some searches might also be sensitive to other decay channels. However, we expect the sensitivity will be low, and it would not change the picture significantly.

\begin{figure}[h!]
   \centering
   \includegraphics[width=1.0\linewidth]{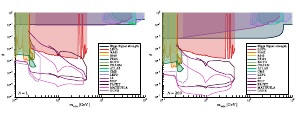}
   \caption{Exclusion Plot for the mixing angle $\theta$ and the mass of the lightest scalar in the low-energy landscape for a broad mass spectrum in the form $[1/\sqrt{N},1]\times m$. The lowest mass region, up to $0.2$ GeV is constrained by fixed target experiments~\cite{CHARM:1985anb, BNL-E949:2009dza, Gorbunov:2021ccu, NA62:2020pwi, KOTO:2020prk, NA62:2021zjw, MicroBooNE:2021sov, MicroBooNE:2022ctm}, above that threshold LHCb searches for $B\to \phi K$ dominate the bounds~\cite{LHCb:2015nkv, LHCb:2016awg, Winkler:2018qyg}. At higher masses we show direct ATLAS, CMS and LEP searches~\cite{L3:1996ome, LEPWorkingGroupforHiggsbosonsearches:2003ing, CMS:2018zvv, ATLAS:2021hbr} and bounds on the Higgs signal strength in dark blue. The sensitivities of future proposals are shown with solid lines~\cite{Curtin:2018mvb, Cerci:2021nlb, Alekhin:2015byh, Berryman:2019dme}. \textbf{Left:} Constraints on a single scalar mixing with the Higgs. \textbf{Right:} Constraints on the lightest scalar from  a sector with a total of $N=200$ scalar singlets. The bounds from inclusive searches become more stringent. Resonant searches for a single scalars,  those by LHCb, ATLAS and CMS can also be enhanced by the presence of multiple scalars within the experimental mass resolution.}
  \label{fig:lightest}
\end{figure}

For reference, in Fig.~\ref{fig:lightest} we show the bounds on the lightest scalar in the spectrum. Even in this case, the presence of a landscape makes a difference, as one can see by comparing the two panels of the Figure. In the left panel we show the bounds on a single scalar mixing with the Higgs. They should be compared to the bounds on the lightest scalar in our landscape that are shown to the right. Indirect constraints (in blue) and inclusive searches (NA62~\cite{NA62:2020pwi, NA62:2021zjw}, E949~\cite{BNL-E949:2009dza}, KOTO~\cite{KOTO:2020prk},  CHARM~\cite{CHARM:1985anb},  MATHUSLA~\cite{Curtin:2018mvb},  FACET~\cite{Cerci:2021nlb},  SHiP~\cite{ Alekhin:2015byh} and DUNE~\cite{Berryman:2019dme}) appear stronger on a landscape than on a single scalar, when displayed in terms of $\theta$, because they are sensitive to multiple new particles (that can be both produced and detected) and we chose to define $\theta$ as the mixing angle for a single particle. 
Some collider searches for resonant signals also get stronger as we increase the number of scalars, since at large $N$ more than one new particle can fall within the mass resolution of the experiment. This is particularly evident for the L3~\cite{L3:1996ome} bound that in the case of a landscape gets stronger at lower masses where the absolute mass splittings between the scalars become smaller.

To summmarize, current experimental probes of new singlets mixing with the Higgs are effective on the few lightest scalars in the landscape, but fail to detect the large multiplicities that would reveal the presence of many vacua. This is summarized in the upper left panel of Fig.~\ref{fig:money}, where we show the branching ratio of the new scalars to two SM particles. Each point is a new scalar and we have generated randomly a sample spectrum from a uniform distribution in $m^2$. For $\lambda^\prime/\lambda \lesssim 10^{-6}$ all new scalars decay mostly to the SM and are detectable with existing searches. As we increase $\lambda^\prime$ past this threshold the branching ratio to the SM quickly goes to zero for all, but a handful of scalars near the bottom of the spectrum. Existing direct searches are essentially insensitive to high-multiplicity decays, as shown in the other panels of the Figure.

From this discussion it emerges that it would be interesting to target these peculiar final states with large multiplicity, but relatively small total energy. In the following we first briefly discuss the production of the new scalars at the LHC and then describe in more detail the anatomy of the scalars' cascade decays.

\subsection{Landscape Production at Colliders}
\label{sec:production}
The collider production of new scalars mixing with the Higgs has been studied extensively (see, e.g., Refs.~\cite{Clarke:2013aya, Ferber:2023iso}). Here we mainly summarize known results and their application to our models. It is convenient to divide the discussion into two mass ranges.

The first one is $m_{\phi}\geq m_B-m_K$, where $m_B \simeq 5$~GeV is the mass of the lightest neutral $B$-meson and $m_K$ of the lightest neutral $K$-meson . In this region gluon fusion is the dominant production process~\cite{Haisch:2016hzu}. We rescale existing results~\cite{Haisch:2016hzu, LHCHiggsCrossSectionWorkingGroup:2016ypw} by the mixing angle between the Higgs and a single new scalar in Eq.~\eqref{mixingangle}.
In this mass range, whenever $m_\phi < m_h/2$, we also study the production of new scalars via on-shell decays of the Higgs boson, even if this channel has a smaller production cross section, because it leads to much longer decay chains and more energetic decay products in the final state. Additionally, reconstructing the Higgs mass can reduce the SM backgrounds (although at high multiplicities invariant mass constraints are not as effective as at low multiplicities). The Higgs width into the new scalars is given in Eq.~\eqref{eq:Hphiphi}. We leave to future work the exploration of other production mechanisms such as $\phi$-strahlung, $\phi V$ and $\overline{t}t\phi$.

\begin{figure}[!t]
    \centering
    \includegraphics[width=1.0\linewidth]{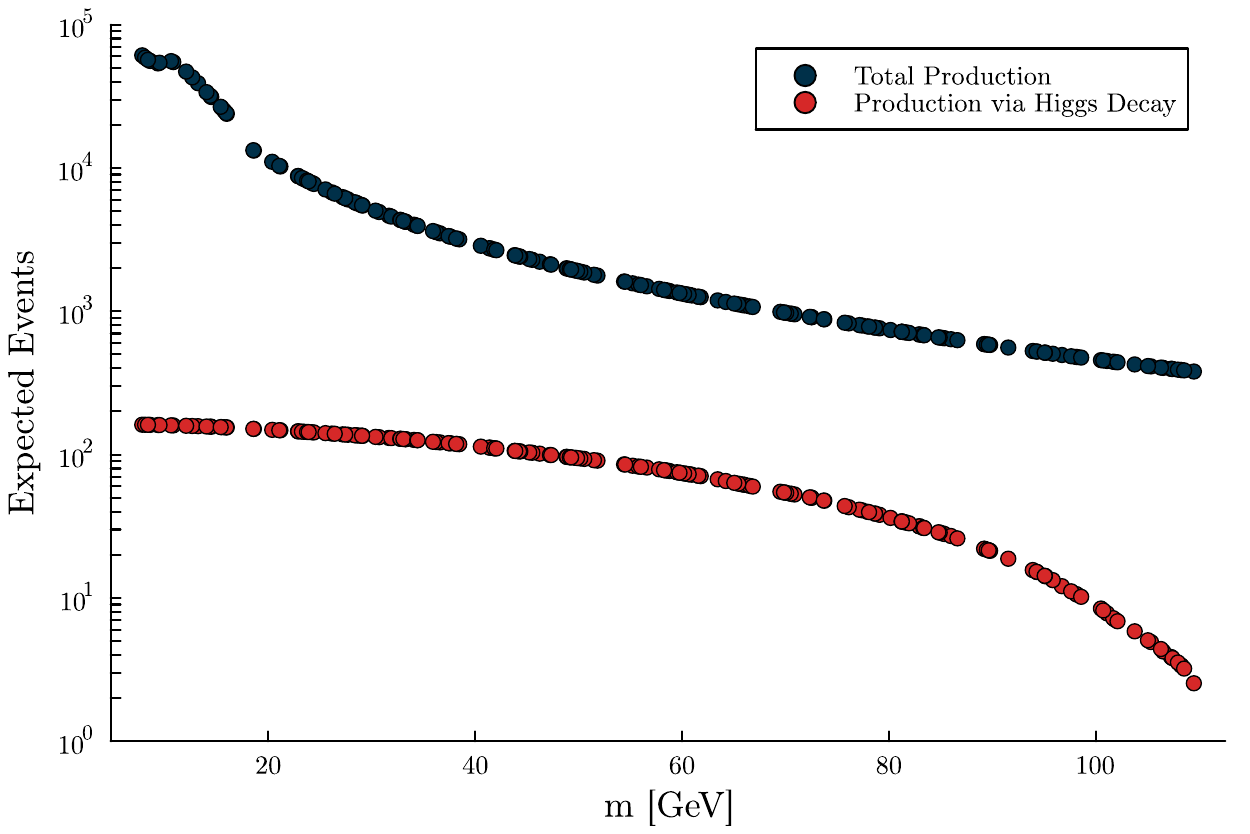}
    \caption{Total number of expected events at the LHC with $\mathcal{L}= 3\times \textrm{ab}^{-1}$ of integrated luminosity for direct $\phi$ production (blue) and number of events from Higgs decays (red). As an example we used a scenario with $\theta = 10^{-3}$, near the boundary of the LHCb sensitivity to a single scalar decaying predominantly to two SM particles, but well below other collider constraints. Each point shows the cross section for a new scalar in a sample mass spectrum.} \label{fig:Production}
\end{figure}

The second regime that we consider is $m_{\phi}< m_B - m_K$. In this case, B-mesons can decay into $\phi$'s. The production cross section via B-mesons is roughly $10^6$ times larger than the one via gluon fusion. The existence of this threshold strongly affects the length of decay chains for mass spectra that cross it, as discussed in the next Section.

We show an illustrative example of the production cross section in Fig.~\ref{fig:Production}, where we plot the total number of expected events for direct $\phi$ production and for production via Higgs decays. The total number of expected events can be sizable at low masses, but these events are challenging to detect due to the small total energy and, even smaller, energy per particle in the final state.

\begin{figure}[h!]
    \centering
    \includegraphics[width=1.0\linewidth]{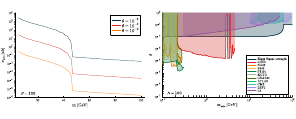}
    \caption{\textbf{Left:} Total production cross section of $N=100$ landscape scalars for a mass distribution $[1/\sqrt{N},1]\times m$. The sudden drop in the production rates at around $m > 45$ GeV, comes from the fact that the lightest scalar can not be produced through $B$ decays.   \textbf{Right:} Constraints on the mixing angle $\theta$ for the lightest scalar in the spectrum of mass $m_{\rm min}$. The lowest mass region, up to $0.2$ GeV is constrained by fixed target experiments~\cite{CHARM:1985anb, BNL-E949:2009dza, Gorbunov:2021ccu, NA62:2020pwi, KOTO:2020prk, NA62:2021zjw, MicroBooNE:2021sov, MicroBooNE:2022ctm}, above that threshold LHCb searches for $B\to \phi K$ dominate the bounds~\cite{LHCb:2015nkv, LHCb:2016awg, Winkler:2018qyg}. At higher masses, we show direct ATLAS, CMS, and LEP searches~\cite{L3:1996ome, LEPWorkingGroupforHiggsbosonsearches:2003ing, CMS:2018zvv, ATLAS:2021hbr} and bounds on the Higgs signal strength in dark blue. 
}
    \label{fig:ProductionII}
\end{figure}

In Fig.~\ref{fig:ProductionII} we do a similar exercise, showing the production cross section of $N=100$ scalars as a function of the typical mass scale $m$ in the spectrum $m_{\phi}=[1/\sqrt{N}, 1]\times m$. In the left panel, we display three cross sections for three different values of the mixing angle $\theta$. In the right panel, we show current constraints on the lightest scalar in the spectrum, appropriately rescaled to take into account the presence of the other 99 scalars, as explained in the previous Section. We see that the new scalars are copiously produced at the LHC, with cross sections as large as $100$~pb, in regions that are still experimentally unconstrained. As stated before the challenge lies in detecting their many-body, but low-energy final states, as we detail in the following.

  \subsection{Cascade Decays}

The typical length of the new scalars' decay chains varies greatly in the parameter space of our toy models. It depends mostly on the mass range of the new scalars. The main signatures can range from traditional two-body final states to long cascades with tens of particles in the detector. Long low-energy cascades are generically expected in the models of the landscape described in Section~\ref{sec:generic}, with the longest cascades taking place in models where the mass distribution spans the largest interval (i.e. $m\times[1,1/\sqrt{N}]$). The specific model described in Section~\ref{sec:joint} leads instead to more standard signatures with two particles in the final state, due to the strong bound on the cross-coupling between the new scalars in Eq.~\eqref{eq:lambda_bound}. The parameters that affect most strongly the collider phenomenology of our models are 
\begin{enumerate}
    \item The typical mass scale of the scalars.

    \item The number of scalars $N$.
    \item The ratio $\lambda^\prime/\lambda$ between the coupling within the landscape and the one to the SM. 
\end{enumerate}
All three parameters affect both production and decays of the new scalars. Before getting into the details, it is useful to go through a broad qualitative overview of the impact of these three quantities on the scalars' phenomenology, starting with the scalars' masses.

\begin{figure}[p]
    \centering   \includegraphics[width=1.0\linewidth]{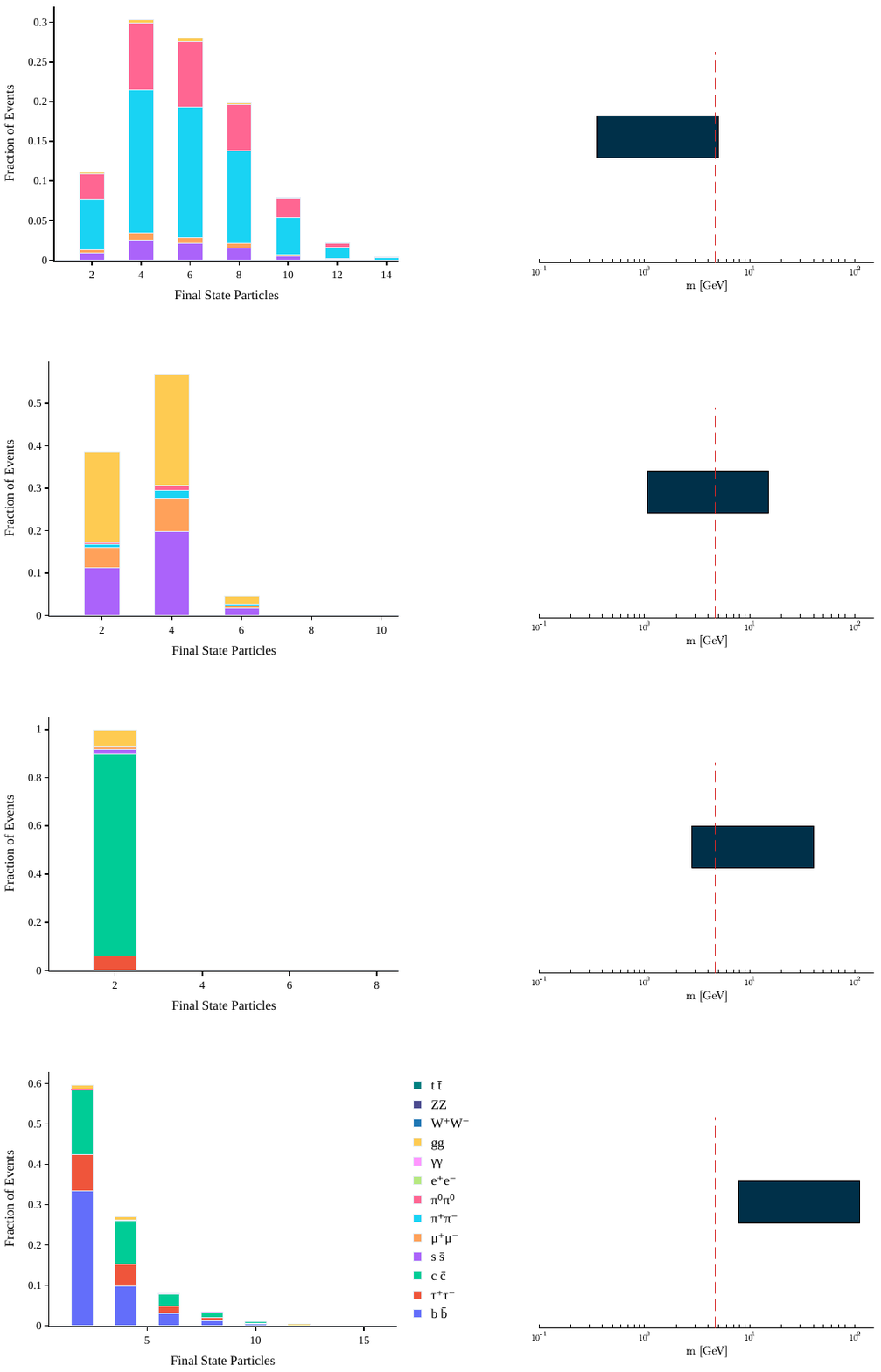}
    \caption{\textbf{Left:} Final state particles from scalars produced via gluon fusion and $B$-meson decays for a scenario with $N=200$ and $\lambda^\prime/\lambda = 1$. \textbf{Right:} Range of the mass distribution of the new scalars relevant to the left panel.}  
    \label{fig:Bth}
\end{figure}

The lightest $B$-meson threshold $m_B \simeq 5$~GeV strongly affects the length of the cascades, as shown in Fig.~\ref{fig:Bth}. If all the scalars are below or above $m_B$, high-multiplicity final states become important, while they can be negligible for scalar mass distributions spanning both sides of the threshold. A lighter scalar that can be produced from $B$ meson decays has a cross section roughly $10^6$ times larger than a heavier scalar that can only be produced through gluon fusion. Having the lightest scalar below  this threshold generically leads to shorter cascades if we average over all the scalars that are produced.  This is due to the fact that the lightest scalars, which can only decay to the SM due to kinematics, are dominantly produced.

A wider mass distribution gives two effects. The larger mass splittings between the scalars favor longer cascades, opening up more decay channels for the heavier scalars. At the same time, lighter scalars are produced with much larger cross sections. 
Typically, our benchmark, where the scalars are distributed in the range  $[1/\sqrt{N}, 1]\times m$ gives longer cascades than the case where both boundaries of the mass distribution are of the same order. Of course, most of the lighter scalars can be produced through B decay if it is kinematically allowed. These scalars will decay to SM final states without going through a long cascade.

The total number of scalars $N$ affects the overall size of the mixing between the SM and the landscape, $\theta\sim 1/\sqrt{N}$, and thus the total number of new physics events expected at colliders. $N$ also impacts the typical length of cascade decays. The median length increases with $N$, but the rate at which it increases depends strongly on the mass distribution, as detailed below. However the typical cascade length is not monotonic in $N$ because of the $B$-threshold. Increasing $N$ can broaden the mass distribution if we assume the mass spectrum is in the range $[1/\sqrt{N}, 1]\times m$. If this makes the lightest scale below the B decay threshold,  the cascade length strongly decreases.

Finally, $\lambda'/\lambda$ has an obvious impact on the length of cascades. As we decrease the ratio the scalars will go more and more into two-body final states, decaying directly to the SM via their mixing with the Higgs in Eq.s~\eqref{eq:V1}, \eqref{eq:V2} and \eqref{eq:Vsimplified}. The total production cross section also depends on the overall value of $\lambda$, since $\theta\propto \lambda$. Interestingly, the transition between long cascades and two-body final states is not at $\lambda^\prime/\lambda \simeq 1$, but at a much lower value. This emerges immediately from the decay widths of the scalars. Even when $\lambda^\prime/\lambda \ll 1$ decays into other scalars can dominate over decays to the SM, since 
\be
\frac{\Gamma_{\phi \to {\rm BSM}}}{\Gamma_{\phi\to {\rm SM}}} \sim \left\{ \begin{array}{c}\left(\frac{\lambda^\prime}{\lambda \, g_{\rm SM}}\right)^2\left(\frac{m_h}{m_\phi}\right)^2\quad m_\phi < m_h\\\left(\frac{\lambda^\prime}{\lambda \, g_{\rm SM}}\right)^2\left(\frac{m_\phi}{m_h}\right)^2\quad m_\phi > m_h\end{array}\right. \, . 
\ee
At fixed values of $\lambda^\prime/\lambda$ lighter scalars have longer cascades because $g_{\rm SM}$ is a Higgs coupling to SM particles and the relevant coupling gets smaller and smaller at low mass. For instance a $200$~GeV scalar can decay to $WW$, but a 0.5~GeV scalar will have to decay through the strange or muon Yukawa couplings or radiative couplings to gluons and photons.

\begin{figure}[h!]
    \centering
\includegraphics[width=1.0\linewidth]{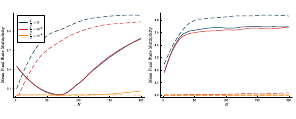}
    \caption{Mean multiplicity of final state particles from new scalars' decays, as a function of the number of scalars $N$. Solid lines correspond to a mass scale $m = 20$~GeV while dashed lines to $m=200$~GeV. The \textbf{left} panel shows a broad mass spectrum $[1/\sqrt{N},1]\times m$. The \textbf{right} panel shows a narrow spectrum $[1/2,2]\times m$.}
    \label{fig:CvsN}
\end{figure}

After this broad overview, we can give some quantitative results on the decays of the new scalars that confirm the qualitative intuition that we have just developed. In the left panel of  Fig.~\ref{fig:CvsN} we see two effects. If increasing $N$ shifts the mass distribution below the $m_B$ threshold, the cascade length can decrease with $N$ (solid lines in the left panel). If it does not (dashed lines everywhere and all lines in the right panel), the average cascade length initially grows linearly with $N$ and then saturates, when the finite phase space becomes important. Note that if phase space was never important, at very large $N$, the cascade length would grow as $\sqrt{N}$, following the arguments in~\cite{DAgnolo:2019cio}. 

\begin{figure}[!h]
    \centering  \includegraphics[width=0.6\linewidth]{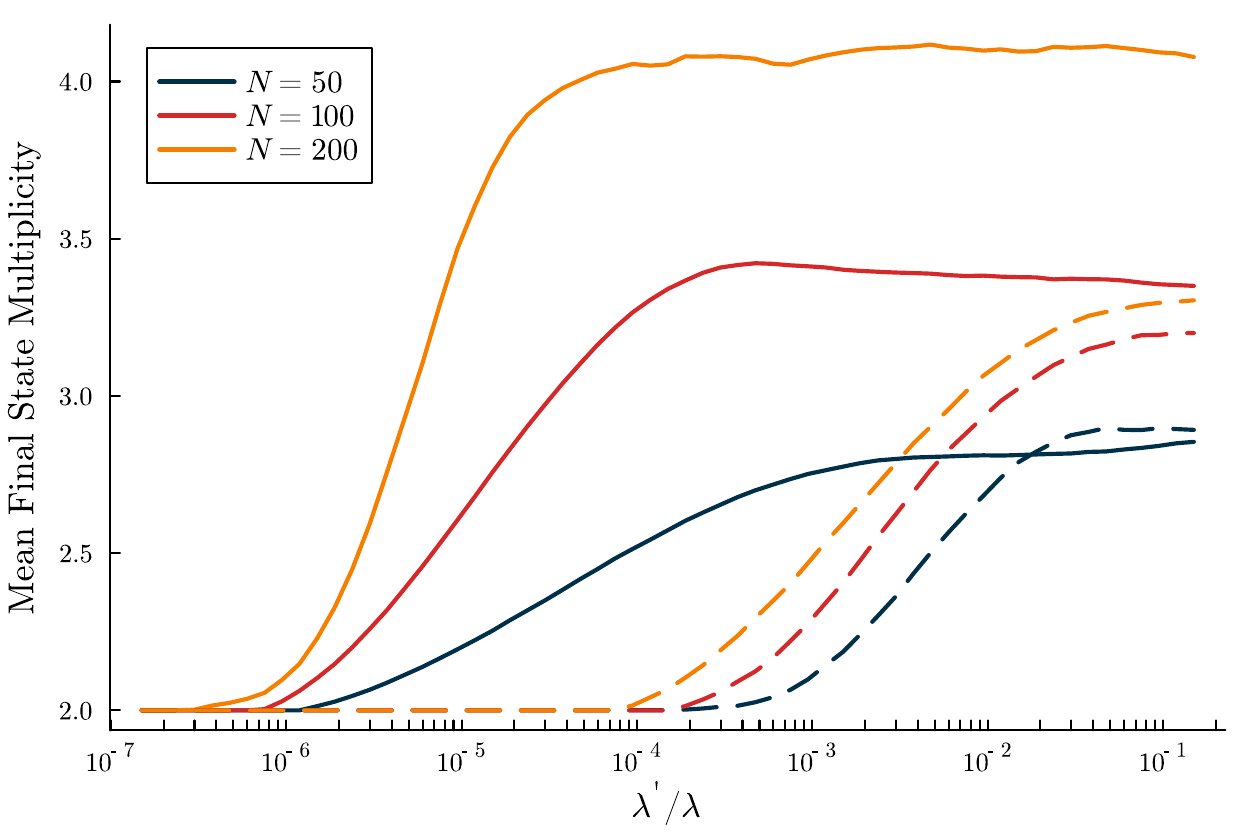}
    \caption{Mean multiplicity of final state particles from new scalars' decays, as a function of $\lambda^\prime/\lambda$ for a broad mass spectrum $[1/\sqrt{N},1]\times m$, with $m=10\textrm{ GeV}$ (solid lines) and $m=200\textrm{ GeV}$ (dashed lines). We show three different choices for the total number of scalars $N$.}.
    \label{fig:Cvslambda}
\end{figure}

For a narrow mass spectrum (i.e., both boundaries of the spectrum of the same order in $N$), the cascade length saturates for some $N=N_*$ as shown in the right panel of Fig.~\ref{fig:CvsN}. This is due to the large occupation of phase space, as anticipated above. As we increase $N$ in a fixed mass range, the scalars get closer and closer to each other so that most decays become kinematically forbidden and the decay length ceases to grow. 
At a fixed value of $\lambda^\prime/\lambda$ lighter scalars (solid lines) have longer cascades compared to heavier ones (dashed lines) because of the smaller couplings to the SM that they inherit from the Higgs, as displayed in Fig.~\ref{fig:Cvslambda}. More broadly, Fig.~\ref{fig:Cvslambda} shows that the cascade length saturates (i.e., decays within the landscape sector become dominant) already for $\lambda^\prime/\lambda \ll 1$, with the precise value depending on $N$ and the mass scale of the scalars.

\begin{figure}[h!]
    \centering
      \includegraphics[width=1.0\linewidth]{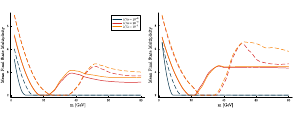}
    \caption{\textbf{Left:} Mean multiplicity of final state particles from new scalars' decays, as a function of the scalar mass scale $m$ for a broad spectrum ($[1/\sqrt{N}, 1]\times m$). Solid lines show $N =50$ while dashed lines $N=100$. \textbf{Right:} The 10\% quantile of events with the highest multiplicity for the same parameters. We display three choices of $\lambda^\prime/\lambda$.}  
    \label{fig:Cvsm}
\end{figure}

In Fig.~\ref{fig:Cvsm} we show the cascade length as a function of $m$ for broad mass spectra. The most prominent feature of the Figure is again the dip in length caused by the $B$ threshold. At low masses we can have very long cascades, then the cascade length decreases when the heaviest states move above the threshold and become a negligible component of the total cross section. The cascade length starts increasing again when most states move past the threshold. After reaching a maximum when all scalars are above the lightest $B$-meson threshold, the cascade length slowly decreases again due to the larger Higgs couplings to the SM that become relevant to the new scalars' decays.

\begin{figure}[h!]
    \centering
      \includegraphics[width=1.0\linewidth]{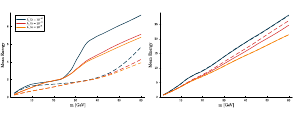}
    \caption{Median energy of final state particles in the scalars' cascade decays for a scenario with $N=50$ (solid lines) and $N=200$ (dashed lines) scalars. On the \textbf{left} the mass distribution is $[1/\sqrt{N}, 1]\times m$, while on the \textbf{right} we consider a narrow spectrum  $[1/2, 2]\times m$. Different colors represent different choices of $\lambda^\prime/\lambda$.}
    \label{fig:Energy}
\end{figure}

Finally, in Fig.~\ref{fig:Energy} we show the median energy of the final state particles for a broad mass spectrum (left panel) and a narrow one (right panel). We see immediately that a traditional high-$p_T$ search for the decay products of our landscapes would fail, especially in the case of a broad spectrum, and similarly, the total energy in the event (determined by $m$) is not enough for a high $S_T$ search. Even targeting the high multiplicities in the final state might be challenging. On the one hand, because of the reduction in cross section that one pays when targeting heavier scalars that decay to higher multiplicity final states, and on the other, because final state particles become extremely soft. The median energy in Fig.~\ref{fig:Energy} is quite challenging to trigger on, for both broad (left panel) and narrow (right panel) mass spectra, even if it contains an important contribution from two-body final states. If our final states contained mostly leptons, narrow mass spectra would be simple to trigger on, as shown in the right panel of the Figure, but this is not the case because the new scalars decay mostly via mixing with the Higgs. Representative final states for different choices of the mass scale are shown in Fig.~\ref{fig:Bth}. In light of this discussion it is worth to consider final states from Higgs decays, as discussed in the next subsection.

\begin{figure}[h!]
    \centering
      \includegraphics[width=0.6\linewidth]{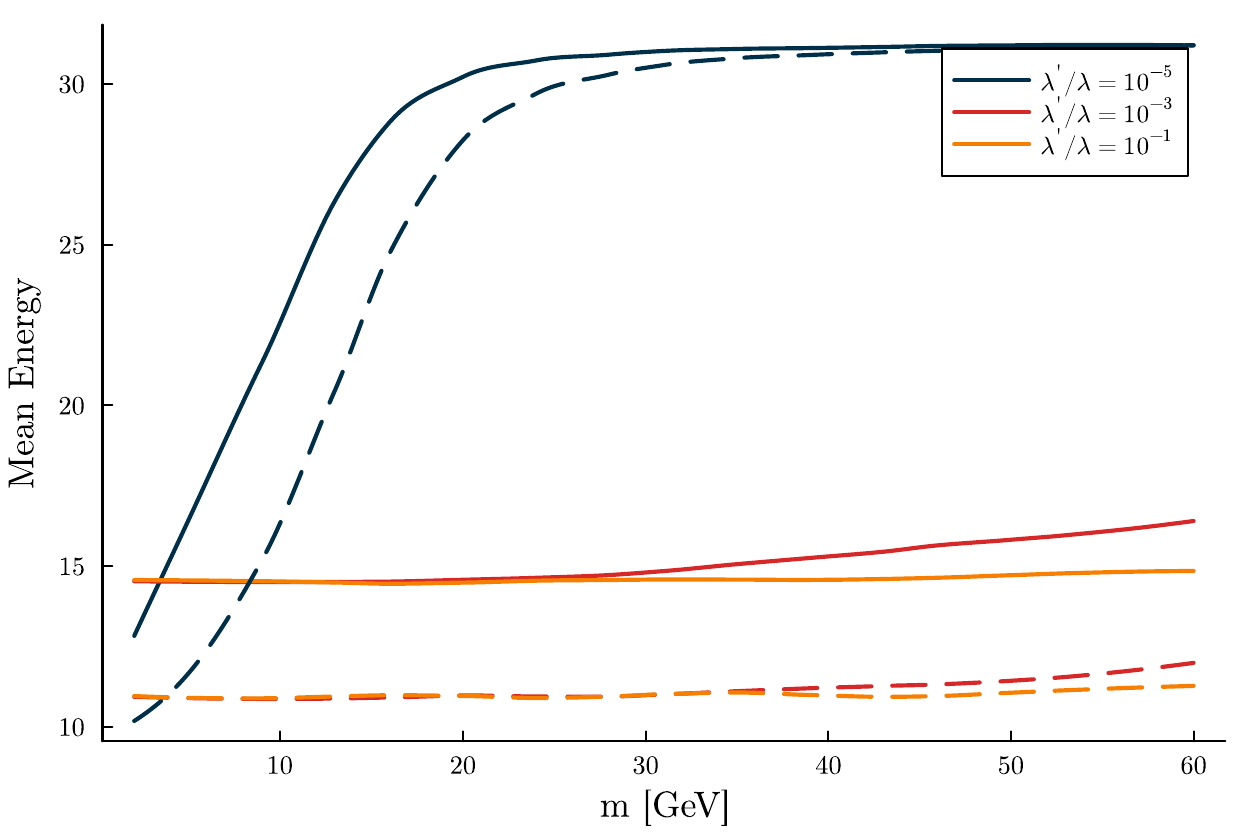}
    \caption{Median energy of the final state particles from the new scalars' decays for scalars produced via Higgs decays. We show $N=50$ (solid lines) and $N=200$ (dashed lines). On the \textbf{left} The mass distribution is $[1/\sqrt{N}, 1]\times m$, while it is  $[1/2, 2]\times m$ on the \textbf{right}. The colored lines represent different choices of $\lambda^\prime/\lambda$.}
    \label{fig:EnergyH}
\end{figure}

\begin{figure}[p]
    \centering   \includegraphics[width=1.0\linewidth]{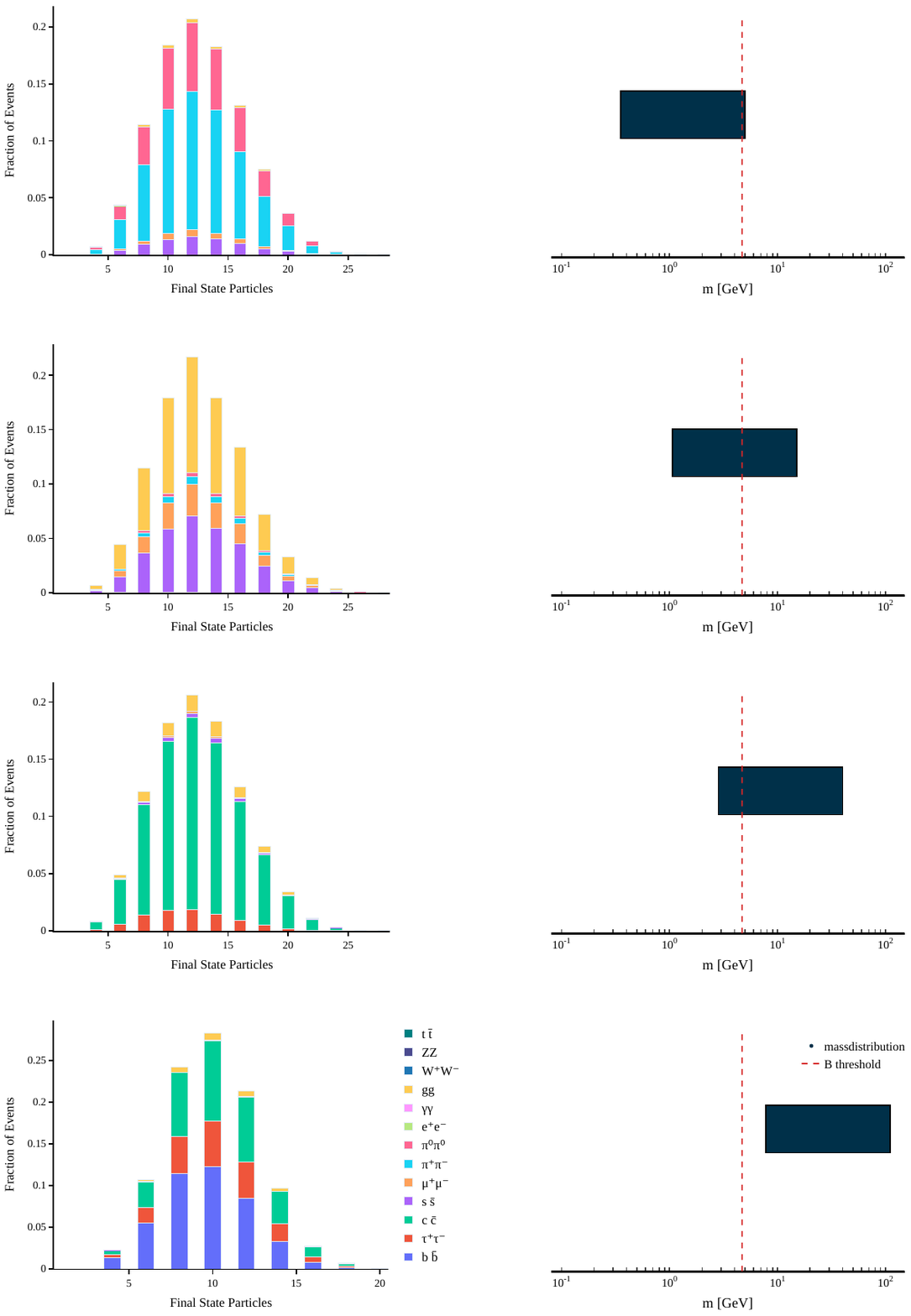}
    \caption{\textbf{Left:} Final state particles from scalars produced via Higgs decays for a scenario with $N=200$ and $\lambda^\prime/\lambda = 1$. \textbf{Right:} The blue box shows the range of the mass distribution of the new scalars relevant to the histogram to its left.}  
    \label{fig:BtH}
\end{figure}

\subsubsection{Higgs Decays}

Since the discovery of the Higgs boson, its exotic decays have been the focus of numerous search efforts (for a broad overview see~\cite{Curtin:2013fra, Cepeda:2021rql}). However, the signal discussed here, in connection with the landscape, is qualitatively distinct. There are significant differences in the composition of the final states and their kinematics, which we will discuss in detail below. In addition, the signal spreads among many different final states, with each one only having a tiny branching ratio. This certainly 
introduces new challenges in identifying this signal efficiently. 

Although it shares certain features with the previously discussed signals of landscape scalars, the signal arising from Higgs decays exhibits important differences. 
The results in the previous Section suggest that landscapes could be challenging to detect at colliders, even if they are in the right energy range. The small total energy in the final state might make them elusive even for the most modern background rejection and triggering techniques. This suggest exploring final states that are rarer, but allow for raising trigger thresholds and offer more handles to reduce SM backgrounds. Higgs bosons decays is an obvious candidate in this category. We have already shown in Fig.~\ref{fig:Production} the much smaller production cross section in this channel compared to gluon fusion and $B$ decays. However, in Fig.~\ref{fig:BtH} we show that cascades from Higgs decays are quite spectacular, with a median number of particles in the final state around 12, regardless of the typical mass scale of the new scalars. The increased cascade length compared to the direct production of the new scalars is due to two effects. First, we always have two landscape scalars after the first stage of the Higgs decay, rather than one in the direct production from gluon fusion. In addition, the scalars are produced almost democratically, as shown in Fig.~\ref{fig:Production}.  Hence, heavier scalars with longer decay chains are produced more often than in single production, where the pdfs of the proton strongly suppress the production of heavy scalars relatively to lighter ones.

Another important quality of this production mechanism is that the median energy of the final state particles is much higher compared to direct $\phi$ production, as shown in Fig.~\ref{fig:EnergyH} (to be compared with the left panel of Fig.~\ref{fig:Energy}, relevant to $\phi$ production). Since the new scalars come from Higgs decays, the dominant energy scale in the process is $m_h$ , and the final state particles have typical energies independent of $m_\phi$. The only case in which we see a large variation of the median energy as a function of $m_\phi$ is for $\lambda^\prime/\lambda \lesssim 10^{-6}$. In this case, the new scalars have only two-body decays when $m$ become larger than 20~GeV (see for instance, Fig.~\ref{fig:Cvsm}), while below they can have longer cascades leading to smaller energies of the final state particles.

A detailed comparison, through a collider analysis, between the Higgs decays and direct $\phi$ production modes goes beyond the scope of this work, but we hope that this qualitative comparison will motivate the LHC collaborations to study both options and also alternative production mechanisms.

\section{Conclusion}\label{sec:conclusion}
In this work, we have studied the collider signatures of a low energy sector of the Multiverse. Discovering the Multiverse would radically alter the way we think about the cosmological constant, the electroweak hierarchy problem, and what is fundamental in a theory of Nature. Therefore, we consider the possibility that a sector of the landscape of vacua that constitutes the Multiverse is within experimental reach. We mainly study the mass range $0.1\;{\rm GeV}\lesssim m_\phi \lesssim 200$~GeV since it has large production cross sections and offers ample detection opportunities without being already excluded by traditional high-$p_T$ searches at the LHC or inclusive searches at beam dump experiments.

In Section~\ref{sec:generic}, we construct a series of toy models that describe generic low-energy quantum field theory landscapes. In Section~\ref{sec:joint}, we construct an explicit model that explains both the CC and the Higgs mass squared through Weinberg's argument and a low energy landscape. In Section~\ref{sec:pheno}, we have examined the generic signatures of landscapes at colliders, through the models in Section~\ref{sec:generic} and~\ref{sec:joint}, and found a rich phenomenology that spans several different possibilities. We could simply have a large number of copies of a singlet scalar decaying predominantly to two SM particles, or models dominated by spectacular cascades that would fill LHC detectors with tens of final state particles. We find that both are equally natural and arise in generic regions of parameter space.

As a consequence, much of the parameter space of these models is still relatively unconstrained by colliders. This is not due to their feeble couplings to the SM. On the contrary their total production cross section can be sizeable and even beyond that of SM electroweak processes that have already been measured at the LHC. The challenge lies in detecting high-multiplicity, but low-energy final states that can easily get lost in the QCD background. 

We also highlighted the production of landscape scalars in  exotic Higgs decays as an interesting benchmark and a possible discovery channel. In particular, the signal features long decay chains and more energetic final state particles, potentially offering better reach even though the total production rate is lower than the gluon fusion channel. 

Even if the likelihood of a low energy sector of the landscape is unknown, a discovery can trigger a real shift of perspective in fundamental physics. If we can find evidence for the Multiverse, we will never think about fine-tuning problems in the same way. These include the size of the observable universe and the mass of most know particles. We hope that the experimental reader will be motivated to pursue this high-risk/high-gain endeavor, apply data-scouting techniques that have already been developed~\cite{CMS:2024zhe, Boveia:2703715, Benson:2019752} and invent new strategies to detect the high-multiplicity, but low-energy final states described in Section~\ref{sec:pheno}.

\acknowledgments

We would like to thank Nima Arkani-Hamed for inspiration and discussions during the early stages of the project.  RTD and ME acknowledge ANR grant ANR-23-CE31-0024 EUHiggs for partial support. This research was supported in part by grant NSF PHY-2309135 to the Kavli Institute for Theoretical Physics (KITP). LTW is supported by the Department of Energy grant E-SC0009924.

\bibliography{refs_clean}
\bibliographystyle{JHEP}
\end{document}